\documentclass[3p,,preprint,12pt,onecolumn]{elsarticle}

\usepackage{lineno}
\usepackage{graphicx}
\usepackage{extarrows}
\usepackage{epstopdf}
\usepackage{threeparttable}
\usepackage{multirow}
\usepackage{float}
\usepackage{booktabs}
\usepackage{tabularx,,etoolbox}
\usepackage[dvipsnames]{xcolor}
\usepackage{ulem}

\newlength{\toprulewidth}
\patchcmd{\toprule}
  {\heavyrulewidth}{\toprulewidth}
  {}{}

\patchcmd{\bottomrule}
  {\heavyrulewidth}{\toprulewidth}
  {}{}

\begin{document}

\begin{frontmatter}

\title{An Iterative Machine-Learning Framework for RANS Turbulence Modeling}

\author[BUAAadd]{Weishuo Liu}
\ead{liuweishuo@buaa.edu.cn}
\author[STFCadd]{Jian Fang\corref{cortextcontent}}
\ead{jian.fang@stfc.ac.uk}\cortext[cortextcontent]{Corresponding author.}
\author[STFCadd]{Stefano Rolfo}
\ead{stefano.rolfo@stfc.ac.uk}
\author[STFCadd]{Charles Moulinec}
\ead{charles.moulinec@stfc.ac.uk}
\author[STFCadd]{David R Emerson}
\ead{david.emerson@stfc.ac.uk}
    
\address[BUAAadd]{School of Energy and Power Engineering\unskip, 
    Beihang University\unskip, 37 Xueyuan Road, Haidian District\unskip, Beijing\unskip, 100191\unskip, China}
  	
\address[STFCadd]{Scientific Computing Department\unskip, 
    Science and Technology Facilities Council\unskip, Daresbury Laboratory\unskip, Keckwick Lane\unskip, Daresbury,
Warrington\unskip, WA4 4AD\unskip, UK}

\begin{abstract}
Machine-learning (ML) techniques provide a new and encouraging perspective for constructing turbulence models
for Reynolds-averaged Navier--Stokes (RANS) simulations. 
In this study, an iterative ML-RANS computational framework is proposed that combines an ML algorithm with transport equations
of a conventional turbulence model. This framework maintains a consistent
procedure for obtaining the input features of an ML model in both the training and predicting stages, ensuring a built-in reproducibility.
The effective form of the closure term is discussed to determine suitable target variables
for the ML algorithm, and the multi-valued problem of existing constitutive theory is studied
to establish a proper regression system for ML algorithms.
The developed ML model is trained under a cross-case strategy with data
from turbulent channel flows at three Reynolds numbers and \textit{a posteriori} simulations of
channel flows show that the framework is able to predict both the mean flow field and turbulent variables
accurately. Interpolation tests for the channel flow show the proposed framework can reliably predict flow 
features that lie between the minimum and maximum Reynolds numbers associated with the training data.
A further test related to the flow over periodic hills also demonstrates a better result
than a traditional turbulence model, indicating a promising predictive capability of the developed
ML model for separated flow even though the model is only trained with planar channel flow data.
\end{abstract}


\end{frontmatter}

\section{Introduction}
The Reynolds-averaged Navier--Stokes (RANS) equations have been widely used for decades in fluid engineering 
and will continue to play an essential role because of their reduced computational cost~\cite{RN66, RN65}. However,
averaging the Navier--Stokes equations introduces new Reynolds stress terms which creates a closure
problem. Over many years, a large number of turbulence models have been
developed~\cite{RN81, RN72, RN63, RN98, RN75, RN82, RN67, RN73} with a physics-driven modeling process,
which normally refers to making hypotheses based on physical intuition, constructing differential or 
algebraic mathematical models, parameterizing the models and calibrating the coefficients. 
The advantage of these physics-driven models is that they can capture the major turbulence transport characteristics, and can be successfully applied to a broad range of engineering applications. However the turbulence model remains the primary
source of uncertainty in RANS simulations~\cite{liu2008,villalpando2011,roy2006review}
because real engineering applications normally deviate from the ideal conditions used to construct these models.

Recently, machine learning (ML), which is known to be an efficient tool for dealing with complex regression
systems~\cite{RN85}, has been able to shed light on the RANS closure problem by providing an alternative data-driven 
method to establish the mapping relation between the closure term and the mean flow field~\cite{RN38}. 
Recent studies have demonstrated the feasibility of the application of
ML algorithms in turbulence modeling utilizing data from high-fidelity simulations~\cite{RN38, RN36}.

In general, a turbulence model is constructed with a combination of transport equations and an
algebraic constitutive law to calculate the Reynolds stress tensor~\cite{RN82, RN89}, such as the two-equation models
using a first-order constitutive relation (i.e., the Boussinesq hypothesis)~\cite{RN72, RN63, RN98, RN67, RN73}.
In terms of applying ML into transport equations, Parish \& Duraisamy~\cite{RN41} and Holland~\textit{et al.}~\cite{RN55}
developed a field inversion technique to find the spatial distribution of the ML target variables. This
strategy, combined with an ML technique such as a neural network (NN), was used to construct a
functional form of correction coefficients in the turbulence transport equations. The methodology was
applied and tested for the Spalart-Allmaras ($S$--$A$) model~\cite{RN31,duraisamy2015new,RN32} and the transitional $k$--$\omega$
model~\cite{duraisamy2015new,RN35}, where successful reproductions of the training cases were achieved. 
More recently, Zhu~\textit{et al.}~\cite{zhu2019machine}
explored the possibility to substitute the partial differential equation (PDE) in the $S$--$A$ model with an algebraic
mapping in NN, demonstrating an advanced fitting ability of the ML model.

As for the constitutive law, existing studies mainly followed the general effective-viscosity hypothesis and the
mathematical form originally proposed by Pope~\cite{RN12}. 
Under this form, the dimensionless deviatoric Reynolds stress tensor is expressed as a function of dimensionless 
mean strain tensor and rotation tensor, which can be finally expressed as a linear combination of a series of 
linearly independent tensors under the theory of tensor analysis~\cite{RN94} and matrix polynomials~\cite{RN95, RN96}. Each 
combination coefficient can be expressed as a function of the invariants of the tensors. Therefore, seeking the functional form of the coefficients and their calibration remains the major problem in modeling 
the constitutive relations. Weatheritt \& Sandberg~\cite{RN57,weatheritt2017development} worked on a novel 
parametric modeling approach, where they utilized
a gene expression program to search for an optimal analytical formula for coefficients of the tensorial combination.
Favorable algebraic expressions were found for separated flows~\cite{RN57} and duct
flows~\cite{weatheritt2017development} respectively, and the application of their model trained with data from large-eddy simulation of a low-pressure turbine wake flow showed improvements against isotropic models~\cite{akolekar2019development}.

Constitutive laws can also be estimated by a non-parametric ML model. 
Ling~\textit{et al.}~\cite{RN29} firstly introduced a specified structure with embedded tensorial invariance to an NN~\cite{RN28}. Wang~\textit{et al.}~\cite{RN60} and
Wu~\textit{et al.}~\cite{RN59} extended the constitutive expression with two more tensors and adopted the
random forest algorithm to select the most relevant features. Both studies provided valuable modeling
methodologies and showed improved predictions of anisotropic flows against linear constitutive models.
However, the prediction yielded a one-step augmentation instead of an iteratively converged simulation.
The closure term was predicted based on the mean flow field of a RANS simulation and was frozen after the
prediction. The final result was obtained by further running the RANS equations with the frozen closure term.
This means that the results of their studies still could be seen as intermediate flow fields instead of
the dynamical fixed point of the ML-integrated PDE system. As the closure term was not frozen but dynamically
updated with the simulation, the simulation could deviate from the result presented in their studies.
Therefore, the design of the computational process and internal technical details need to be further
investigated and developed.

In Ling's framework~\cite{RN29, RN28}, there is an inconsistency between the turbulence quantities (e.g., $k$
and $\varepsilon$) used in the training and predicting stages. In the training process, statistically averaged
turbulence quantities (e.g., $k$, $\varepsilon$) from the high-fidelity database were directly used to obtain the
training data. However, in the prediction process, the ML model was actually giving predictions based on
turbulence quantities estimated by traditional turbulence models. As a consequence, structural uncertainties
were introduced into the framework by the different procedures to obtain turbulence quantities between
training and predicting stages. 
Another important issue is that the primary hypothesis of the general effective-viscosity theory of Pope~\cite{RN12} used by
Ling~\textit{et al.}~\cite{RN29, RN28} appeared to be a multi-valued function (see Sec.~\ref{Sec:2B}). 
This issue was also reported by Sotgiu~\textit{et al.}~\cite{sotgiu2019} when applying ML methods into the 
mathematical form of the general effective-viscosity theory. The multi-valued problem could bring great difficulties to deterministic ML regression models, 
as it would possibly destroy the continuity of the machine-learned functional mapping, which would in turn lead 
to a divergence state of the differential system.
In contrast, the physics-informed ML framework of Wang~\textit{et al.}~\cite{RN60} and Wu~\textit{et al.}~\cite{RN59}
introduced more independent tensors into the constitutive relations, as they derived an eddy-viscosity-based
decomposition form of the closure term to ensure consistency between a model's \textit{a priori}
and \textit{a posteriori} performance~\cite{RN50}. However, the computational architecture was compromised
into an open-loop process. During the training process of the ML model, conventional RANS mean flow features
were used as input and closure terms from high-fidelity simulations were used as a training target. Therefore,
the mapping relation was actually constructed between an inaccurate mean flow and an accurate closure term,
which brought difficulties in predicting global flow phenomena such as the area of a separation region.
The possible reason for the loss of convergence could have been the choice of the random forest algorithm,
which is a sum of simple step functions with an intrinsic absence of continuity and differentiability.

In the present study, an iterative computational framework between the ML model and the RANS solver is constructed, in which the consistency of input features between training and predicting stages is ensured. The eddy-viscosity-based decomposition form of the closure term is adopted, and the multi-valued problem is overcome by expanding the set of input variables. Direct Numerical Simulation (DNS) data of turbulent channel flows are then used to train the ML model with a cross-case strategy, where data at different Reynolds numbers are merged together and fed to the NN. It is demonstrated that the model trained with the channel data not only gives accurate results for channel flows at different Reynolds numbers within the training range, but also presents a better performance against a well-established RANS model in flow with a large separation.

The paper is organized as follows; the methodology is introduced in Sec.~\ref{Sec:Methodology}, with the
structure of the developed ML-RANS framework and a brief discussion on the necessary measurements to take
for reproducibility of the training cases (Sec.~\ref{Sec:Framework}). Next, the two major challenges,
the effective form of the closure term and the multi-valued problem of the general constitutive hypothesis,
are discussed in Sec.~\ref{Sec:2A} and Sec.~\ref{Sec:2B}, respectively. The training results
and the \textit{a posteriori} tests of the developed ML model are shown in Sec.~\ref{Sec:Results},
and conclusions are drawn in Sec.~\ref{Sec:Conclusion}.

\section{\label{Sec:Methodology}Methodology}
In this section, an iterative ML-RANS framework is proposed to derive a consistent procedure to obtain the turbulence quantities used in the training and prediction stages. Within the framework, two significant topics in ML-based turbulence modeling, including the effective form of the closure term and the multi-value problem in existing constitutive relations, are discussed based on previous research and data analysis in a turbulent channel flow.

\subsection{\label{Sec:Framework}Iterative ML-RANS Framework}
The Reynolds averaged Navier-Stokes equations for incompressible flow written
for steady state read
\begin{equation}
    \begin{aligned}
        \nabla &\cdot \boldsymbol{u}=0
        \\
        \boldsymbol{u}\cdot \nabla \boldsymbol{u}-\nu \nabla ^2\boldsymbol{u}+ &\nabla p-\nabla \cdot
\boldsymbol{\tau }-\boldsymbol{f}=0 
    \end{aligned}
    \label{eq:1}
\end{equation}
where $\boldsymbol{u}$ is the mean velocity vector, $\nu$ is the molecular viscosity, $p$ is the mean pressure,
the term $\boldsymbol{\tau }$ is the unclosed Reynolds stress tensor and $\boldsymbol{f}$ is the body force acting
on the continuum.

When the steady-state RANS equations are solved in an iterative
manner~\cite{patankar2018numerical,ferziger2002computational},
the procedure refers to solving the turbulence transport equations and applying a constitutive relation for the definition of the unclosed Reynolds stress tensor. The procedure is shown as follows, in which the subscript $n$ represents the solution for current the
iteration and $n-1$ for the previous iteration.

\begin{enumerate}
    \item The initial mean velocity and pressure fields, namely $\boldsymbol{u}^0$ and $p^0$,
     respectively, are given, as well as the initial turbulence quantities.
    \item The coefficients in the constitutive relation (e.g., eddy-viscosity) are calculated based on the 
    turbulence quantities at the previous iteration. 
    \item The momentum equations with continuity correction including a Poisson's equation for the pressure
     are solved using closure term and mean field ($\boldsymbol{u}^{n-1}$, $p^{n-1}$) from 
    the previous iteration. The mean flow field is then updated to ($\boldsymbol{u}^{n}$, $p^{n}$).
    \item The turbulence transport equations are solved based on the updated mean field
    ($\boldsymbol{u}^{n}$, $p^{n}$).
    \item If the simulation has not converged, go back to Step~2.
\end{enumerate}

\begin{figure}[!ht]
    \centering
    \includegraphics[width=1\textwidth,trim=85 140 85 60,clip]{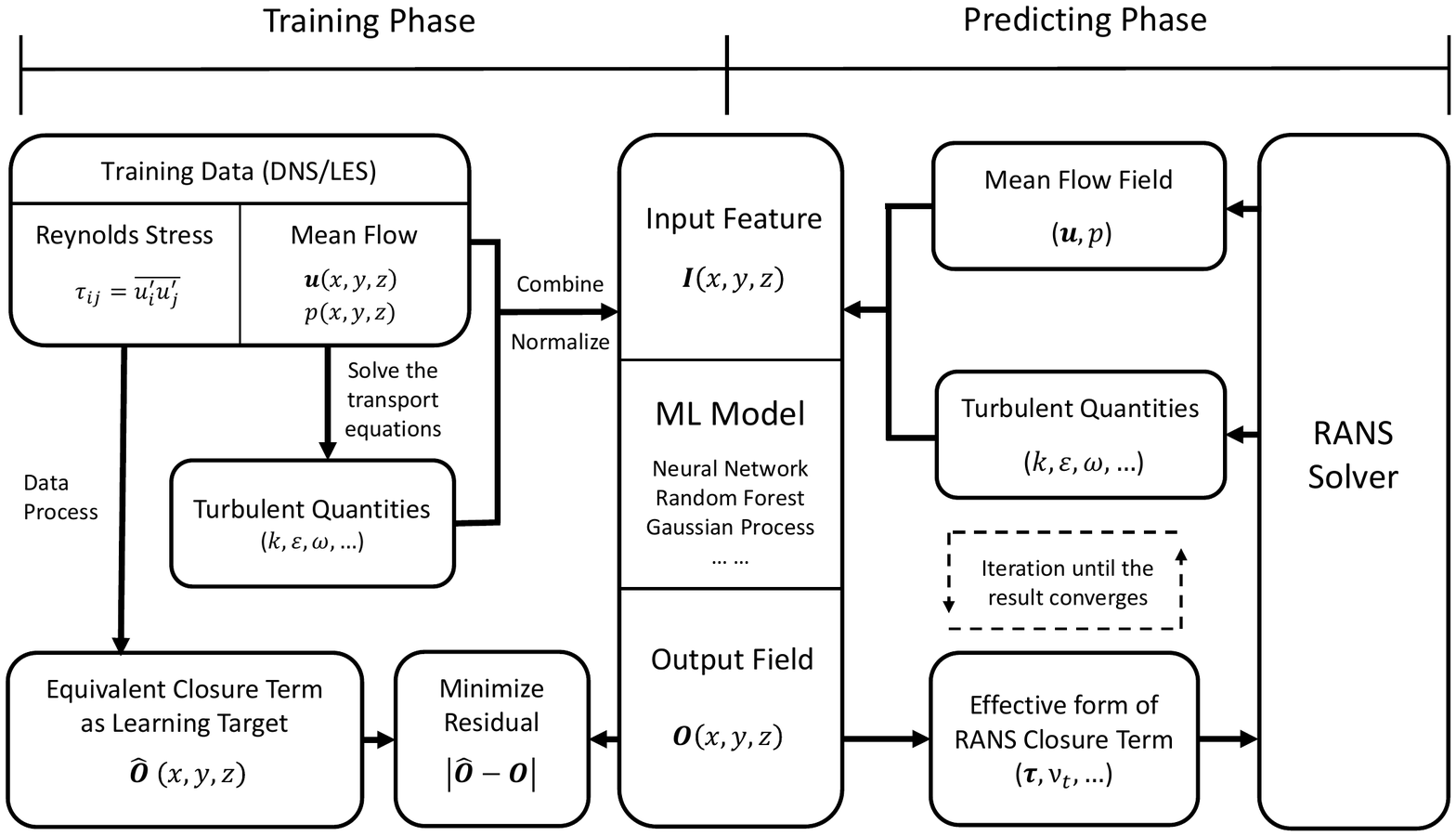}
    \caption{\label{fig:framework}A comprehensive sketch for the iterative ML-RANS framework.}
\end{figure}

The ML regression technique can be
integrated into this computational process at either Step~2 or Step~4 for the calculation of an
effective closure term.
In this study, we implement the ML model at Step~4 to assist the calculation of the closure term. 
Dimensionless functional mapping is embedded into the ML model, where the input features consist of the mean
flow field and transport-equation-estimated turbulence quantities.
The overall process including training and
predicting stage is shown in Fig.~\ref{fig:framework}.

In the training stage, the input feature and target variables for the ML model are obtained from the training data, which are statistics of high-fidelity DNS data, such as, mean velocity field, turbulent kinetic energy and turbulent dissipation rate. It is natural to obtain these turbulence quantities directly from DNS data according to their definitions, such as previous ML-RANS frameworks~\cite{RN29,RN60,RN59}. However, it has to point out that the real values of these turbulence quantities are usually unknown in a general RANS simulation, and their estimations are obtained by solving the transport equations of a turbulence model. For the consistency between training and predicting phases, the turbulence quantities used to compose the input feature during the training phase are also obtained by solving the transport equations of a traditional turbulence model based on the mean flow field of the training DNS data. In the following predicting phase, the input feature can be obtained in the same manner by replacing the DNS mean flow field used in the training phase to the run-time local flow field. 

The input features of the ML model, $\boldsymbol{I}(x,y,z)$, are dimensionless variables 
calculated and normalized by combining the mean flow field and estimated turbulence quantities. The
target variables of the output data for the ML model are processed from the statistically averaged Reynolds
stress of high-fidelity DNS data, and a non-dimensional equivalent form of the closure term is
then calculated as $\boldsymbol{\hat{O}}(x,y,z)$.
The data pairs $\left\{ \boldsymbol{I},\boldsymbol{\hat{O}}\right\}$
are fed into the ML model at the training stage and the training process is finished when the error
between the ML model's output $\boldsymbol{O}(x,y,z)$ from the ML model and the target $\boldsymbol{\hat{O}}(x,y,z)$ is reduced to a certain level. By the end of the training stage, an accurate mapping $\boldsymbol{O}=f(\boldsymbol{I})$
is established to support the calculation of the closure term.

In the prediction stage, the ML model obtained from the above training stage is loaded to a computational fluid dynamics (CFD) solver.
At each iteration step, the ML model receives the input field $\boldsymbol{I}(x,y,z)$ calculated by the
CFD solver and returns the output field $\boldsymbol{O}(x,y,z)$ for the calculation of the closure
term. The CFD solver then receives the output from the ML model, updates the closure term, and solves the
RANS equations to get the mean flow field and turbulent quantities used for the next step. This iterative
loop keeps going until the residuals converge to the given tolerance. 

\begin{figure}[!ht]
    \centering
    \includegraphics[width=1\textwidth,trim=50 260 50 100,clip]{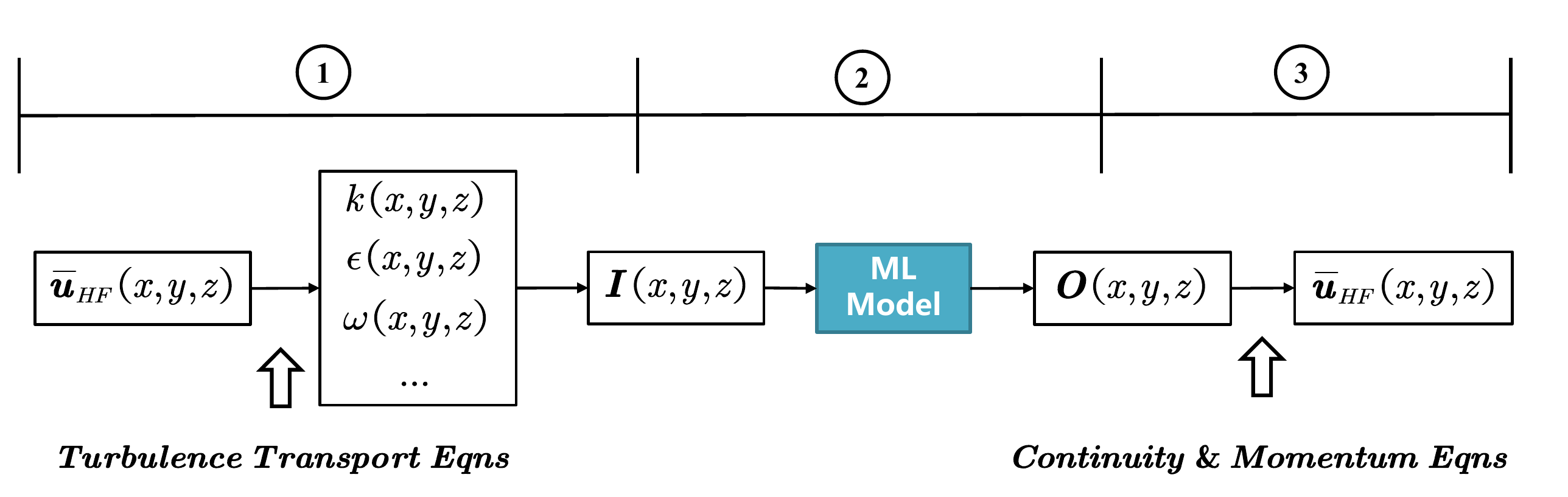}%
    \caption{\label{fig:frame_pre}Prediction phase in iterative ML-RANS framework given HF mean flow as
initial field. The design of the framework should satisfy the following requirements: (1)~the procedure's
consistency of obtaining the input vector $\boldsymbol{I}$. (2)~the functional mapping
relation $\boldsymbol{O}=f(\boldsymbol{I})$ can be established successfully. (3)~the effective
equivalent form of closure term mapped into the flow domain ensures RANS equations converge to the correct mean flow field.}
\end{figure}

With the high-fidelity mean flow field mapped as the initial field, the solution of the
ML-integrated RANS solver should be the same as the high-fidelity results, so that a built-in reproducibility
of the training cases can be achieved. By analyzing the prediction phase of this computational framework shown
in Fig.~\ref{fig:frame_pre}, if the mean flow field from high-fidelity simulations is mapped as the initial field, the
structural design could ensure the same input vectors $\boldsymbol{I}$ for the ML model (steps in Circle 1),
since the input vector $\boldsymbol{I}$ for training data is obtained in the same procedure. To further
obtain the same flow field as the high-fidelity simulations, two significant challenges also need to be
discussed and ensured
\begin{itemize}
\item[-] What equivalent form of closure term (as the learning target for ML models) is able to lead the
RANS solver to converge to the mean flow field of the high-fidelity simulations (Circled 3 in
Fig.~\ref{fig:frame_pre})?
\item[-] Can a successful functional mapping $\boldsymbol{O}=f(\boldsymbol{I})$ be established
throughout ML algorithms (Circled 2 in Fig.~\ref{fig:frame_pre})? 
\end{itemize}

In the following subsections, the discussions are focused on the aforementioned two challenges, along with
relevant preliminary solutions used in this computational framework. The effective form of the Reynolds stress
tensor is discussed in Sec.~\ref{Sec:2A}, and the initial solution to build a one-to-one mapping relation for 
ML models are shown in Sec.~\ref{Sec:2B}.

\subsection{\label{Sec:2A}Effective Form of the RANS Closure Term}

The objective of this section is to present an optimal way to incorporate an expression equivalent to the high-fidelity Reynolds stress tensor, $\boldsymbol{\tau}_h$, into the RANS equations  in order to compute a solution in close proximity to the high-fidelity mean flow field $(\boldsymbol{u}_{h}, p_{h})$, in which the subscript, ${h}$, stands for the variables from high-fidelity simulations.
The solution of the RANS equations should be very close to the high-fidelity
mean flow field $(\boldsymbol{u}_{h},p_{h})$ if the high-fidelity Reynolds stress field $\boldsymbol{\tau}_h$ is mapped 
into the flow domain. However, 
Wu~\textit{et al.}~\cite{RN50} have reported that using an explicit form for the statistically
averaged Reynolds stress tensor as a source term in the RANS equations does not naturally guarantee a correct 
\textit{a posteriori} solution for the mean velocity and pressure field. The RANS equations 
commonly used in steady-state incompressible solvers, 
shown in Eq.~(\ref{eq:2}), would face an ill-conditioned problem 
when the Reynolds stress tensor field is directly added to the 
right-hand side as part of the equation

\begin{equation}
    \left(\boldsymbol{u}^{n-1}\cdot \boldsymbol{\nabla} - \boldsymbol{\nabla} \cdot 
\nu \boldsymbol{\nabla}\right)\boldsymbol{u}^n=
-\boldsymbol{\nabla} p^{n-1}+\boldsymbol{\nabla} \cdot \boldsymbol{\tau}^{n-1} + \boldsymbol{f}^{n-1},
    \label{eq:2}
\end{equation}
in which, the left hand side of the equation, $\mathcal{L}(\ )=\left(\boldsymbol{u}^{n-1}\cdot \boldsymbol{\nabla}
- \boldsymbol{\nabla} \cdot \nu \boldsymbol{\nabla}\right)$, is treated implicitly as the linear operator
on the unknown, $\boldsymbol{u}^n$, and the right hand side, $-\boldsymbol{\nabla} p^{n-1}+\boldsymbol{\nabla} 
\cdot \boldsymbol{\tau}^{n-1} + \boldsymbol{f}^{n-1}$, is treated explicitly as known variables from 
the previous iteration step.

If the high-fidelity mean flow fields are interpolated into the flow domain with $(\boldsymbol{u}^{n-1},p^{n-1})=(\boldsymbol{u}_h,p_h)$
and $\boldsymbol{\tau}^{n-1}=\boldsymbol{\tau}_h$, the expected solution should be $(\boldsymbol{u}^{n},p^{n})\approx
(\boldsymbol{u}_h,p_h)$. 
However, with the problem being ill-conditioned, the error in the solution $\|\delta\boldsymbol{ u}^n\|$
can be extremely sensitive to the statistical errors in the closure
$\|\boldsymbol{\nabla} \cdot \delta \boldsymbol{\tau}^{n-1}\|$ ($\|\cdot\|$ denotes the norm of the discretized vector).
Results of existing research focusing on channel flows have shown that the velocity field at steady-state could deviate
from the high-fidelity solution by over 35\% at a high Reynolds number~\cite{RN47},
and similarly, identical Reynolds stress fields obtained from different data sources 
could result in vastly different solutions in terms of mean velocity~\cite{RN50}.
To circumenvent the issue yielded by the explicit approach not being suitable, 
Wu~\textit{et al.}~\cite{RN50} proposed to compute the eddy-viscosity field from 
the high-fidelity database and interpolate
it into the flow domain as part of the left-hand side of the equation, which could 
improve the conditioning of the linearized momentum equation. To achieve this goal, 
the closure term, $\boldsymbol{\tau}$, is decomposed into its linear and nonlinear parts, as shown in Eq.~(\ref{eq:3}),

\begin{equation}
\boldsymbol{\tau} =2\nu _t\boldsymbol{S}+\boldsymbol{\tau }_{\bot} \label{eq:3} 
\end{equation}
where the linear part, $2\nu _t\boldsymbol{S}$, is the production of turbulence eddy viscosity, $\nu _t$, and mean strain rate tensor, $\boldsymbol{S}$. The non-linear, $\boldsymbol{\tau }_{\bot}$, is the remnant of $\boldsymbol{\tau}$ after being approximated by $2\nu _t\boldsymbol{S}$. Equation ~(\ref{eq:3}) is further split into its implicit and explicit parts in the computation, as shown in Eq.~(\ref{eq:4}).

\begin{equation}
\boldsymbol{\tau} =\nu_t\boldsymbol{\nabla}\boldsymbol{u}^n + \nu_t\boldsymbol{\nabla}^T\boldsymbol{u}^{n-1}
                  + \boldsymbol{\tau }_{\bot}^{n-1} \label{eq:4}
\end{equation}
where the superscripts $ {n}$ and $ {n-1}$ stand for the velocity field of the current and previous steps, respectively, and $\boldsymbol{\nabla}^T\boldsymbol{u}$ is the transpose of the velocity gradient matrix. With the implicit part passed to the left-hand side (see Eq.~(\ref{eq:4})), the linearized momentum equation with improved conditioning reads 

\begin{equation}
    \left(\boldsymbol{u}^{n-1}\cdot \boldsymbol{\nabla} -\boldsymbol{\nabla} \cdot (\nu+\nu_{t})
\boldsymbol{\nabla} \right) \boldsymbol{u}^n =-\boldsymbol{\nabla} p^{n-1}+\boldsymbol{\nabla}
\cdot \boldsymbol{\tau }_{\bot}^{n-1}+\boldsymbol{f}^{n-1}+\boldsymbol{\nabla}\cdot\
\nu_t\boldsymbol{\nabla}^T\boldsymbol{u}^{n-1}
\label{eq:5}
\end{equation}
As a result of the better-conditioned momentum equation obtained by the implicit treatement of the linear part,
small errors in the eddy-viscosity closure will not be amplified 
and propagated to the solution, and the solution of the new equations can be very close to the mean flow 
of the high-fidelity simulations.

Currently, the estimation of the optimal eddy-viscosity is computed by minimizing the
discrepancy between the linear part of the stress and the DNS Reynolds stress data, shown in Eq.~(\ref{eq:6}).

\begin{align}
    \nu_{t}^{m} & =\underset{\nu_{t}}{\arg \min }\left\|\boldsymbol{\tau}_h-2 \nu_{t} \boldsymbol{S}_h\right\|
\label{eq:6}\\
    \nu_{t}^{m} & = \left| \left. \frac{\tau_{h_{ij}}S_{h_{ij}}}{2S_{h_{ij}}S_{h_{ij}}} \right| \right. \label{eq:7}
\end{align}
Equation~(\ref{eq:7}) illustrates the point-wise least squares approximation to estimate eddy-viscosity
from the high-fidelity database.
It should be mentioned that this estimation might encounter numerical discontinuity where
the misalignment of the principal axis of $\boldsymbol{S}_h$ and $\boldsymbol{\tau}_h$ changes rapidly, for example,
at the boundary of a re-circulation zone in a flow with separation.
For this reason, a universal solution for the estimation of $\nu_t$ from a high-fidelity database still needs to be further investigated.

In the present study, the turbulent channel flow is taken as an example
 to illustrate the overall process of the computational
framework and adopt Eq.~(\ref{eq:7}) to process eddy-viscosity from DNS databases. Note that in the case of planar
channel flow with homogeneous streamwise and spanwise directions, only the Reynolds shear stress, $\tau_{12}$,
has a contribution to the RANS equation (see Eq.~(\ref{eq:1})). Therefore, the linear part of the Reynolds stress
could give the totally equivalent effect to the momentum equations, so the turbulence eddy-viscosity, $\nu_t$, becomes
the only term that needs to be modeled under this type of flow. In the present study, the eddy-viscosity scaled by the molecular viscosity, $\nu_t^*=\nu_t^m/\nu$, is taken as the learning target for the ML regression model, and the
improved conditioning equations should ensure that if a correct eddy-viscosity field is predicted by the
ML model, the RANS solver will return a correct mean velocity.

\subsection{\label{Sec:2B}Constructing a Proper ML Regression System based on the Constitutive Theory}

First-order RANS turbulence models are built from a constitutive relation between the Reynolds stress and
the velocity gradients. The simplest one is based on the Boussinesq approximation while the most general
form was proposed by Pope~\cite{RN12}.
This constitutive relation relies on a basic fundamental
assumption, which is that the normalized deviatoric tensor of Reynolds stress, $\boldsymbol{b^*}$, is a function of the
normalized strain and rotation tensors, $\boldsymbol{{S}^*}$ and $\boldsymbol{{\varOmega}^*}$,
as presented in Eq.~(\ref{eq:8}),

\begin{equation}
\boldsymbol{b^*}=\boldsymbol{f}\left( \boldsymbol{{S}^*,{\varOmega}^*} \right),
\label{eq:8}
\end{equation}
in which the normalized deviatoric tensor of Reynolds stress is
$\boldsymbol{b^*}=\dfrac{\boldsymbol{\tau}}{k}-\dfrac{2}{3}\,\boldsymbol{E}$,
where $\boldsymbol{E}$ is the identity matrix 
and $k$ is the turbulence kinetic energy. The normalized strain and rotation tensor are respectively
defined as $\boldsymbol{{S}^*}=\dfrac{1}{2}\dfrac{k}{\varepsilon}\left(\boldsymbol{\nabla}\boldsymbol{u}
+\boldsymbol{\nabla}^T\boldsymbol{u} \right)$,
$\boldsymbol{{\varOmega}^*}=\dfrac{1}{2}\dfrac{k}{\varepsilon}
\left(\boldsymbol{\nabla}\boldsymbol{u}-\boldsymbol{\nabla}^T\boldsymbol{u} \right)$, where $\varepsilon$
is the dissipation rate of turbulence kinetic energy.

Under tensor analysis theory (i.e., the Cayley-Hamilton theorem~\cite{RN94}) and the 
matrix polynomial expansion~\cite{RN95,RN96}, Eq.~(\ref{eq:8}) is equivalent to a finite linear combination
of linearly independent tensors, $\boldsymbol{T}^{(m)}$, which can be written as
\begin{align}
	\boldsymbol{b^*} &=\sum_{m=1}^{10}{G^{\left( m \right)}}\boldsymbol{T}^{\left( m \right)}  \label{eq:9}  
\end{align}
where $G^{(m)}$ is the combination coefficient of the independent tensors, $\boldsymbol{T}^{(m)}$.
The full expression of the independent tensors $\boldsymbol{T}^{(m)}$, can be found in~\ref{Ap:A}. 
Under this mathematical expression, the modeling task can be simplified to determine a functional mapping, $g^{(m)}$, between combination coefficients, $G^{(m)}$, and tensor invariants, $\lambda_l$ as,
\begin{equation}
    G^{(m)}= g^{(m)}(\lambda _1,...,\lambda _5 ) \label{eq:10}
\end{equation}
where $\lambda_l$ ($l = 1$ to $5$) is defined as follows:
\begin{equation}
    \lambda _1=tr\left( \boldsymbol{{S}^*}^2 \right) ,  \lambda _2=tr\left( \boldsymbol{{\varOmega}^* }^2 \right) , 
    \lambda _3=tr\left( \boldsymbol{{S}^*}^3 \right) ,  \lambda _4=tr\left( \boldsymbol{{\varOmega}^* }^2\boldsymbol{{S}^*} \right) ,  
    \lambda _5=tr\left( \boldsymbol{{\varOmega}^* }^2\boldsymbol{{S}^*}^2 \right) \label{eq:11}
\end{equation}
and $tr$ is the trace of the tensors.\\ 

The existing turbulence constitutive relations can be treated as the truncated form of Eq.~(\ref{eq:9}).
For example, the linear eddy-viscosity model (i.e., the Boussinesq hypothesis) is based on the first term
of Eq.~(\ref{eq:9}), whereas the quadratic eddy-viscosity model~\cite{craft1996development}
keeps $\boldsymbol{T}^{(1-4)}$ for a better approximation. However, one significant problem in the basic assumption of this constitutive relations (see Eq.~\ref{eq:8}) is that the independent variables (i.e., $\boldsymbol{{S}^*}$ and $\boldsymbol{{\varOmega}^*}$) are not capable of determining the function value of $\boldsymbol{b^*}$ from the DNS data. That is to say, even the most advanced form of the existing constitutive relations is a multi-valued function, where the function
value, $\boldsymbol{b^*}$, might return very different answers even if the same independent variables,
$\boldsymbol{{S}^*}$ and $\boldsymbol{{\varOmega}^*}$, are used as input data.

\begin{figure}[!ht]
\centering
\includegraphics[width=0.8\textwidth,trim=0 5 0 25,clip]{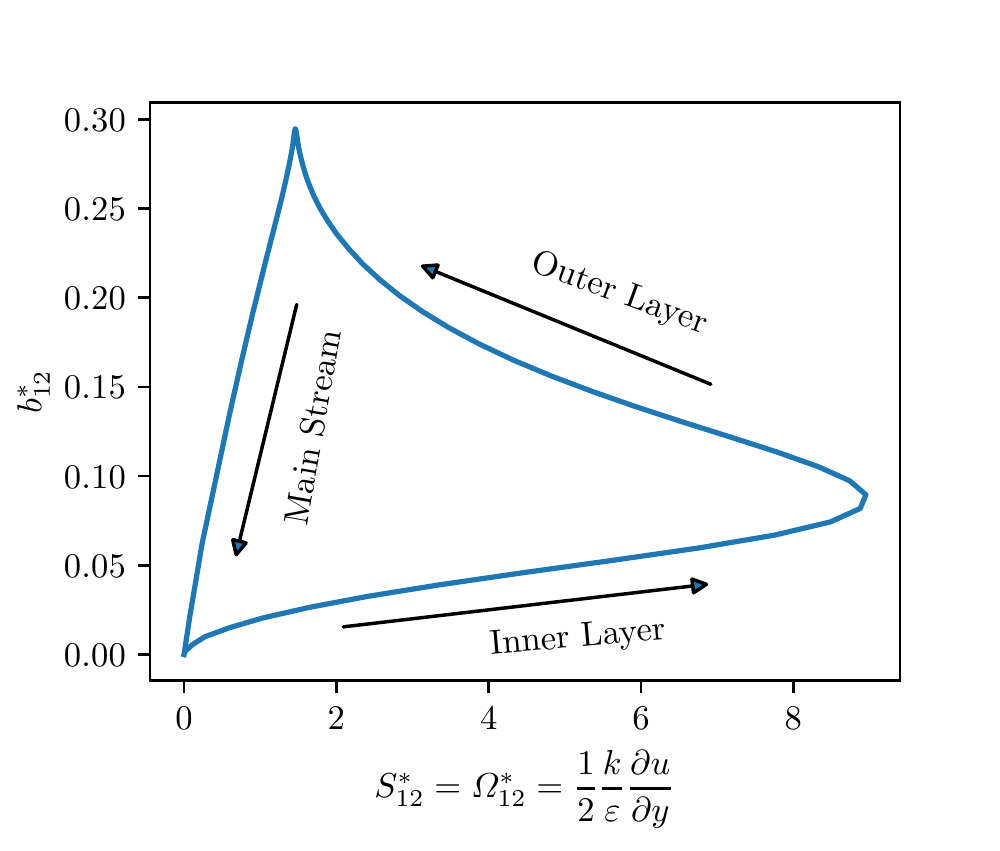}%
\caption{\label{Fig:AS12}Constitutive relation in turbulent channel flow}
\end{figure}

To explain the previous argument, the pattern of how $\boldsymbol{b^*}$ changes with $\boldsymbol{{S}^*}$
and $\boldsymbol{{\varOmega}^*}$ is looked at in a turbulent channel flow using the data of Abe~\textit{et al.}~\cite{RN62} 
at $Re_\tau$ = 180
($Re_\tau = \dfrac{u_\tau h}{\nu}$ with $u_\tau=\sqrt{\nu\dfrac{du}{dy}}$ and $h$ is the half height of channel).
The statistically averaged quantities in the DNS channel case only have
one degree of freedom which changes along the wall-normal direction. Both $\boldsymbol{{S}^*}$
and $\boldsymbol{{\varOmega}^*}$ are solely determined by $\dfrac{k}{\varepsilon}\dfrac{\partial u}{\partial y}$,
as all the other components are zero. The constitutive relation can be visualized
in Fig.~\ref{Fig:AS12}, with
${S^*_{12}}={\varOmega^*_{12}}=\dfrac{1}{2}\dfrac{k}{\varepsilon}\dfrac{\partial u}{\partial y}$
in the $x$-axis and $b^*_{12}=\dfrac{\tau_{12}}{k}$, the only effective component of the
momentum equations, normalized by the turbulent kinetic energy in the $y$-axis.

Figure~\ref{Fig:AS12} shows that the constitutive relation determined by $\boldsymbol{{S}^*}$
(resp. $\boldsymbol{{\varOmega}^*}$)
is not a single-valued function. For example, given the same input
value (e.g., ${S^*_{12}}=4$ in Fig.~\ref{Fig:AS12}), the DNS data returns two different function
values ($b^*_{12} \approx 0.15$ in the outer layer and $b^*_{12} \approx 0.05$ in the inner layer). 
Treating any one of these two values as noise and filtering it out would be inappropriate because
all the different branches of the data have clear physical meanings.
However, for most deterministic ML methods (e.g., NN),
the algorithms are designed under the principal premise that the input-output is a one-to-one mapping
relation~\cite{hassoun1995fundamentals,shinki2001approximation}. Handling such a multi-valued function would make
the ML regression model struggle to minimize the mean square error
for the training data. Considering the example of two data points located on different branches, namely
$P_1(x_0,y_1)$ and $P_2(x_0+\delta x,y_2)$, but where their abscissa only differs by a small number $\delta x$,
the very different output data would lead the ML models to face a tremendous increase from $y_1$ to $y_2$
inside the small $[x_0, x_0+\delta x]$ interval, leading to large oscillations occurring in the
machine-learned mappings~\cite{fang2019neural}, and affecting the differentiability
and stability of the PDE system. Even if the model is perfectly regularized, the mean values
between the different branches are returned to the RANS solver, and considerable structural uncertainties are brought into the PDE system.

To avoid the aforementioned issue and treat the problem as multi-valued,
the set of independent variables needs to be expanded. This leads to
building a proper regression system for ML algorithms. The advantage of the tensorial combination form (Eq.~\ref{eq:9})
is that the Reynolds stress tensor changes accordingly with the tensor bases $\boldsymbol{T}^{(m)}$ without
changing the coefficients $G^{(m)}$. Therefore, such a mathematical form is independent from any rotational
transform of the frame of reference, and adding more rotational invariant scalars $q_j$ to the set of independent
variables for the coefficients $G^{(m)}$ does not change the tensorial expressions. Moreover, in order to
ensure that the model does not depend on any specific types of flow,
the extra scalars $q_j$ should
be defined as a dimensionless combination of local flow variables (e.g., $k$, $\varepsilon$, etc) rather than
macro quantities (e.g., mass flow rate, Reynolds number, angle of attack, etc). Therefore, the functional
form of $G^{(m)}$ can be expanded as,

\begin{equation}
    G^{(m)}=g^{(m)}(\lambda_1,...,\lambda_5,q_1,q_2,...)  \label{eq:12}
\end{equation}

The aim of this work is to obtain a functional expression for the dimensionless eddy-viscosity
in order to overcome the ill-conditioned problem.
By replacing the dimensional definition into Eq.~(\ref{eq:9})
with $\boldsymbol{S}=\dfrac{k}{\varepsilon}\cdot\boldsymbol{{S}^*}$, $\boldsymbol{\tau}
=k\cdot(\boldsymbol{b^*}+\dfrac{2}{3}\boldsymbol{E})$,
the dimensional form of the constitutive equation reads:

\begin{align}
    \boldsymbol{\tau } & =\frac{k^2}{\varepsilon}\left( G^{\left( 1 \right)}\boldsymbol{S} \right)
+k\sum_{m=2}^{10}{G^{\left( m \right)}}\boldsymbol{T}^{\left( m \right)}+\frac{2}{3}k\boldsymbol{E} \label{eq:13}
\end{align}

The first term of the right-hand-side of Eq.~(\ref{eq:13}) is the linear term. By substituting Eq.~(\ref{eq:3}), the functional form of the turbulence eddy-viscosity can be expressed as the coefficient of the linear term,
\begin{equation}
    \nu _{t}=\frac{1}{2}\frac{k^2}{\varepsilon}G^{\left( 1 \right)} \label{eq:14}
\end{equation}

For example, the original $k$--$\varepsilon$ model~\cite{RN72} gives $\nu_t=C_{\mu}\dfrac{k^2}{\varepsilon}$ as a
special constant function for $G^{(1)}$, which is $G^{(1)}=2C_{\mu}$. From Eq.~(\ref{eq:14}), one could derive that the dimensionless
eddy-viscosity $\nu_t^*=\dfrac{\nu_t}{\nu}$, which is the learning target for this study, and can be written as:

\begin{equation}
    \nu _{t}^*=\dfrac{\nu_t}{\nu}=\dfrac{1}{2}\frac{k^2}{\nu\varepsilon}G^{\left( 1 \right)}
=f(\lambda_1,...,\lambda_5,q_1,q_2,...,\frac{k^2}{\nu\varepsilon}) 
    \label{eq:15}
\end{equation}

Equation (\ref{eq:15}) shows that the eddy-viscosity depends on the dimensionless
coefficient $G^{(1)}$ and a special normalization factor ${k^2}/{(\nu\varepsilon)}$, which is the ratio between
the eddy-viscosity predicted by traditional RANS models, ${k^2}/{(\nu\varepsilon)}$, and the molecular viscosity, $\nu$.
Because this factor is coordinate-transformation invariant, it can be included
in the additional set of independent variables, $\boldsymbol{q}$.

Based on our experence, the ML model trained with $\boldsymbol{\lambda}$ in the functional definition 
of $\nu_t^*$ can cause a numerical instability during the iteration process. 
Therefore, the final functional form for the ML task can be written as:
\begin{equation}
    \nu _{t}^{*}=f\left( q_1, q_2, ... \right)
\end{equation}
where the special factor ${k^2}/{(\nu\varepsilon)}$ must be included in the set $\left\{ q_1, q_2, ... \right\}$.
In this way, the multi-valued problem can be avoided in the ML regression system because enough independent
variables for the determination of the local eddy-viscosity are used. 

In this work, the conventional SST $k$--$\omega$ model~\cite{RN98} is adopted 
because of its numerical stability and integrability in the near-wall region. However, the
production of $k$ and $\omega$ shares the same dimension with $\varepsilon$, so it is thus
equivalent to~$\varepsilon={k}/{\omega}$. The dimensionless scalar becomes ${k}/{(\nu\omega)}$,
via the substitution of $\varepsilon$. The full list of the additional independent set of $\boldsymbol{q}$ can be found
in Table~\ref{Table:1}, along with their physical meanings and the adjustment to have a relatively uniform
range for the input features.

\begin{table}[h]
    \caption{List of non-dimensional input features for ML model \label{Table:1}}
    \centering
    \begin{threeparttable}
    \begin{tabular}{clccc}
    \toprule [0.2ex]
    \textbf{Variable} & \textbf{Description}                                                                                                                     & \textbf{Definition}                                                   & \textbf{Normalization} & \textbf{Actual input}                          \\ \hline 
    \\[-8pt]
    $q_1$             & Turbulence intensity                                                                                                                     & $k$                                                                   & $\displaystyle\frac{1}{2}U_iU_i$       & $\displaystyle\frac{25k}{25k+0.5U_iU_i}$                          \\[13pt]
    $q_2$             & Normalized factor                                                                                                         & $\displaystyle\frac{k}{\nu \omega}$                                                & Not applicable\tnote{ c}                      &  $\displaystyle\frac{k}{50\nu \omega}$                      \\[13pt]
    $q_3$             & Wall distance                               & $\displaystyle d$    \tnote{*}                                               & $\displaystyle\frac{\nu}{\sqrt{k}}$    & $\displaystyle\frac{\sqrt{k}d}{50\nu}$                      \\[13pt]
    $q_4$             & \begin{tabular}[c]{@{}l@{}}Cross diffusion \\of $k$ and $\omega$ \end{tabular}  & $\displaystyle\frac{\partial k}{\partial x_i}\displaystyle\frac{\partial \omega}{\partial x_i}$ & $\omega ^3$               & $10\left( \displaystyle\frac{1+680\chi _{k}^{2}}{1+400\chi _{k}^{2}}-1 \right) $\tnote{**}             \\[13pt]
    $q_5$             & \multirow{4}{*}{\begin{tabular}[c]{@{}l@{}}Variables in SST \\model to characterize \\ viscous sublayer and \\turbulent region\end{tabular}} & $\displaystyle\frac{\sqrt{k}}{\omega d}$                                           & Not applicable            & $\displaystyle\frac{5\sqrt{k}}{\omega d}$                    \\[13pt]
    $q_6$             &                                                                                                                                          & $\displaystyle\frac{\nu}{\omega d^2}$                                              & Not applicable            & $\displaystyle\frac{200\nu}{\omega d^2}$   \\[14pt]     
    \bottomrule
    \end{tabular}
    \begin{tablenotes}
    \footnotesize
    \item[ c] Not applicable means the normalization is not necessary.
    \item[ *] $d$ is the distance to the wall.
    \item[**] $\chi _k=\max \left( \displaystyle\frac{1}{\omega ^3}\displaystyle
\frac{\partial k}{\partial x_i}\displaystyle\frac{\partial \omega}{\partial x_i},0 \right) $ 
    \end{tablenotes}
    \end{threeparttable}
\end{table}

\subsection{\label{Sec:numerical}Summary of Methodology and Numerical Platform}

To sum up, with the aim that the RANS solver could reproduce the mean flow results of high-fidelity simulations,
the training stage of the ML-RANS framework introduces conventional turbulence transport equations to process
the high-fidelity mean flow, which ensures a consistent procedure to calculate the input features in the predicting
stage (see Sec.~\ref{Sec:Framework}). Furthermore, the decomposition of the Reynolds stress and the implicit treatment
of the RANS equations ensures that they do not amplify the statistical errors present in the closure term
so that a well-predicted closure term can lead the RANS solver to converge to the correct results
(Sec.~\ref{Sec:2A}). However, the previously defined constitutive relation exhibits a multi-valued problem
when results are compared to DNS data, which proves difficult to be solved by ML algorithms.
Here the set of independent variables is expanded to construct a one-to-one
mapping in order to build a proper regression system for ML algorithms (see Sec.~\ref{Sec:2B}). With the initial
solutions in Sec.~\ref{Sec:2A} and Sec.~\ref{Sec:2B}, the computational steps corresponding to  Circles 2 and 3
in Fig.~\ref{fig:frame_pre} can be fulfilled so that the computational framework provides a built-in
solution which is the same as high-fidelity mean flows. 

As for the numerical platform for CFD, the preparation of the training data refers to interpolating the high-fidelity
mean flow field onto a RANS-designed mesh and running the traditional turbulence transport equations as scalar transport
equations. This part is programmed in the Finite-Volume open-source CFD code, OpenFOAM~\cite{RN61}.
The improved conditioning RANS equations (Eq.~\ref{eq:4}) for the prediction phase are also implemented
with the  SIMPLE~\cite{ferziger2002computational, patankar1983calculation} algorithm to deal with
pressure-velocity coupling.

For the training of the ML models, Google's Tensorflow~\cite{RN84} machine learning framework is adopted for the construction of the NN~\cite{RN84, RN100, RN101, RN68}. The turbulence quantities are extracted from the scalar transport equations, along with the mean flow
variables to assemble the input features (Table~\ref{Table:1}). The NN is trained independently and
loaded through a self-developed Tensorflow-OpenFOAM-Interface (TOI)~\cite{TOIref}, where the interface
feeds the current flow field as inputs to the ML model, and receives the output from the Tensorflow library. 

Based on the discussions above, the calculation process is summarized as follows:

\begin{enumerate}
    \item An accurate mean flow field is provided by DNS, based on which the transport equations of $k$
and $\omega$, for the $k$--$\omega$ SST turbulence model are solved.
    \item The input features for the ML model, $\left\{ q_j \right\}$, are constructed as functions of the mean velocity and turbulent quantities as shown in Table~\ref{Table:1}. 
    \item The learning target (i.e., an accurate eddy-viscosity) for the ML model is obtained from the DNS data
by calculating $\nu _t$ using Eq.~(\ref{eq:7}).
    \item The mapping between $\left\{ q_j \right\}$ and the eddy-viscosity is stored in the ML model after the
training stage. 
    \item The OpenFOAM RANS solver computes the input features based on the current (initial) mean velocity and
pressure field.
    \item The ML model receives the current inputs and returns the eddy-viscosity field to the RANS solver
through TOI.
    \item The RANS solver receives the eddy-viscosity from the ML model and computes the corresponding
velocity field, using Eq.~(\ref{eq:5}).
    \item The process is repeated from Step~5 to Step~7 until the residual reaches the convergence criteria.
\end{enumerate}

\section{Results and Discussion \label{Sec:Results}}  
In this section, an NN with a continuous activation function is trained using the DNS data
from planar turbulent channel flows at low to moderate Reynolds numbers.
The DNS data from Abe~\textit{et al.}~\cite{RN62} and Moser~\textit{et al.}~\cite{RN51} (the latter
is known as the KMM database)
are used as the training and validation datasets, respectively. 
In the training process, the training data are used by the optimization algorithms to adjust weights
in the NN, whereas the validation data are used to monitor and control the training process by adjusting the
hyperparameters such as the structure of the NN and learning rate. 
The newly developed ML-RANS model
is first tested in turbulent channel flows at the same Reynolds number as the training data to verify 
the reproducibility of the training cases, and at Reynolds numbers equally distributed
within the range of training data to assess the interpolation capability. 
A further test in flow over periodic hills~\cite{RN54} is also performed to validate the predictive capability
in a flow regime that deviates from the training cases.
Details of the training dataset and test cases are given in Table~\ref{Table:2}.

\begin{table}[!h]
\caption{ List of training dataset and testing dataset \label{Table:2}}
\centering
\begin{threeparttable}
\begin{tabular*}{\textwidth}{c @{\extracolsep{\fill}} ccc}
\toprule
\multicolumn{2}{c}{\textbf{Training Phase}}                                                         & 
\multicolumn{2}{c}{\textbf{Prediction phase}}
   \\[2pt] \hline \\[-9pt]
\textbf{Training set} & \textbf{Validation set}      & \multicolumn{2}{c}{\textbf{Test case}}   \\
Planar channel~\cite{RN62} & Planar channel~\cite{RN51} & Planar channel & Periodic hills \\ \hline \\[-8pt]
\begin{tabular}[c]{@{}c@{}}$Re_\tau$ = 180\tnote{ a}\\ $Re_\tau$ = 395\\ $Re_\tau$ = 640\end{tabular} &
$Re_\tau$ = 587   & \begin{tabular}[c]{@{}c@{}}$Re_\tau$ = $Re_\tau^{train}$,$Re_\tau^{valid}$\\
$Re_\tau$ = 180,230,...,630\end{tabular} & $Re_H$ = 1400\tnote{ b} \\
\bottomrule          
\end{tabular*}
\begin{tablenotes}
\footnotesize
\item[a] $Re_\tau$ is the Reynolds number based on the mean wall friction velocity and the half-height of a channel.
\item[b] $Re_H$ is the Reynolds number based on the bulk velocity and the height of the hill.
\end{tablenotes}
\end{threeparttable}
\end{table}

\subsection{Training Phase}
There are six raw input variables for the NN, $q_1-q_6$, as listed in Table~\ref{Table:1}, and the output
data is the dimensionless eddy-viscosity calculated by Eq.~(\ref{eq:7})
and scaled by the molecular viscosity, $\nu$. 

\begin{figure}[!ht]
\centering
\includegraphics[width=1\textwidth,trim=0 40 0 0,clip]{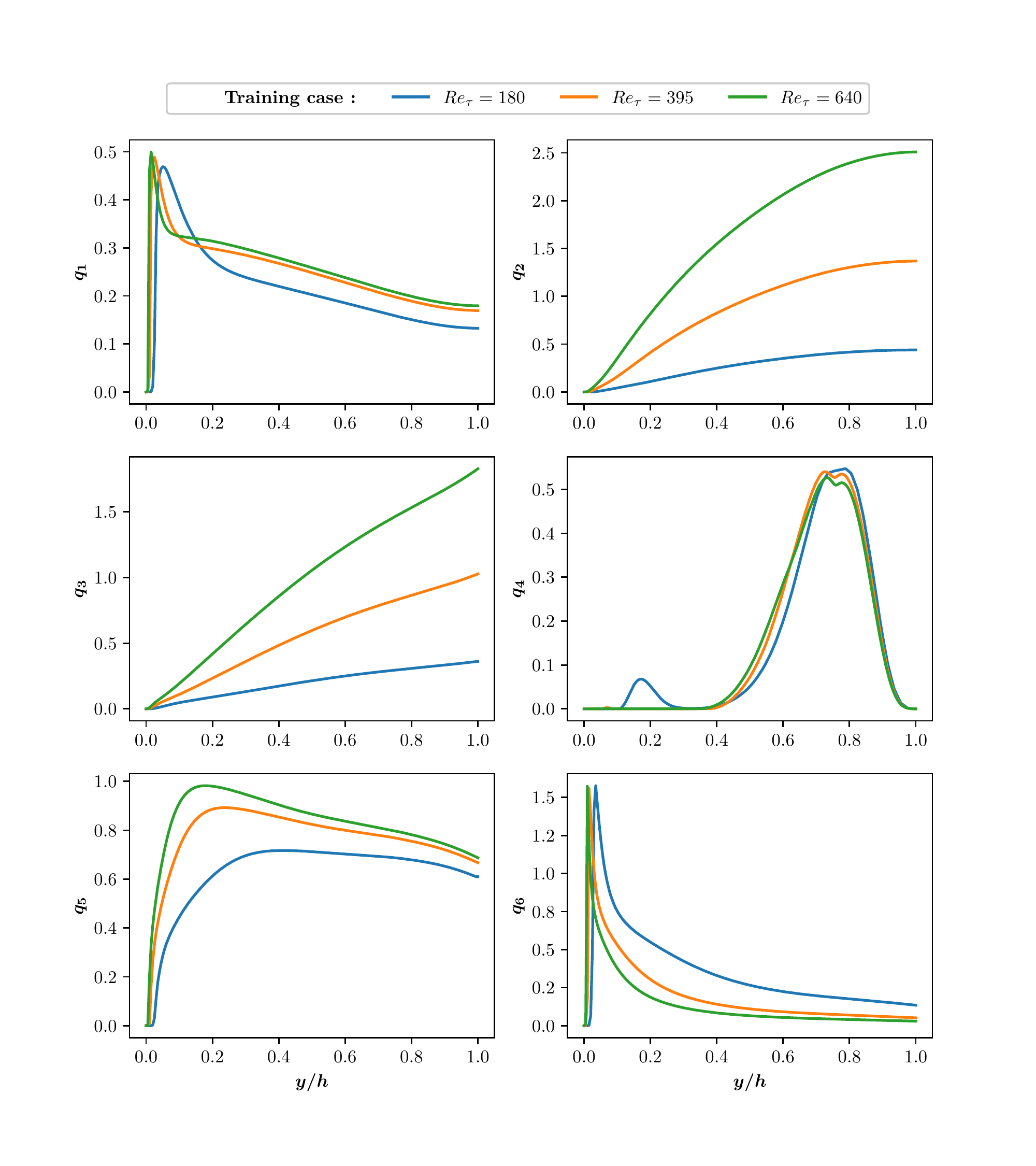}%
\caption{\label{Fig:Inputs_nt}Profiles of the input variables of the NN in channel flows}
\end{figure}

\begin{figure}[!ht]
\centering
\includegraphics[width=1\textwidth,trim=0 15 20 15,clip]{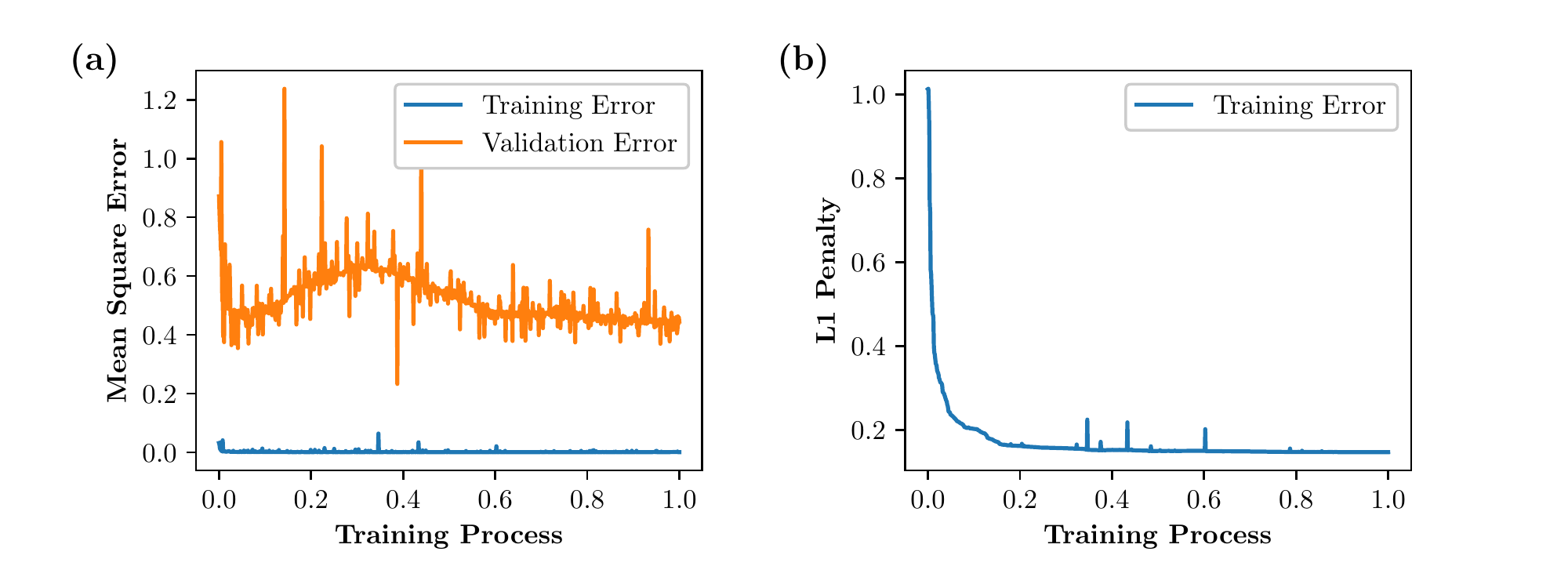}%
\caption{\label{Fig:Errors_nt}Losses in the training process. (a): MSE loss  (b): L1 penalty term. The $x$-axis
is defined as the ratio of current training step and total training steps.}
\end{figure}

The variation of the input profiles, $q_n$, as functions of the wall distance is shown in Fig.~\ref{Fig:Inputs_nt}. 
Note that the intrinsic degree of freedom in this regression system is \textbf{two} because any variable
in the flow can be uniquely determined by the dimensionless distance to the wall, $y^+$, and Reynolds
number of the flow. Fig.~\ref{Fig:Inputs_nt} shows four different patterns for the input parameters, $\left\{ q_n \right\}$. 
Therefore, the degree of freedom of the input is larger
than the degree of freedom of the regression system, so that a one-to-one mapping relation can be
easily established through the ML algorithm. Moreover, we can also observe similar distributions
between $q_2$ and $q_3$, and between $q_1$ and $q_6$ as well. This indicates the six input variables
are redundant for the regression system. Consequently, regularization techniques are needed
to eliminate multicollinearity. In this work, the L1 regularization strategy~\cite{RN102}
is adopted, which shrinks the weight of the NN layers to zero to achieve a better generalization capability. 
The total loss, which is to be minimized in the training of a NN, is defined as,
\begin{equation}
Loss=\dfrac{1}{N}\sum_{k=1}^N{\left( f\left( \boldsymbol{x}_k \right) -y_k \right) ^2}+\lambda \sum_{i,j}^{m,n}{\left| \omega_{ij} \right|} \label{eq:17}
\end{equation}
where \{$\boldsymbol{x}_k$, $y_k$\} are the independent and dependent variables in the training data, and
$f\left( \boldsymbol{x}_k \right)$ is the dependent value predicted by the NN.
$N$ is the number of the data points, $m$ and $n$ are the numbers of layers and nodes in the NN, respectively.
The optimization is applied to both parts of the loss. The first part is the mean square error (MSE) from
the training data, and the second part is the L1 regularization penalty, minimizing the weight of
each layer~\cite{RN102}. The evolution of errors during the training process can be seen in Fig.~\ref{Fig:Errors_nt}.

\begin{figure}[!ht]
    \includegraphics[width=1\textwidth,trim=0 10 20 15,clip]{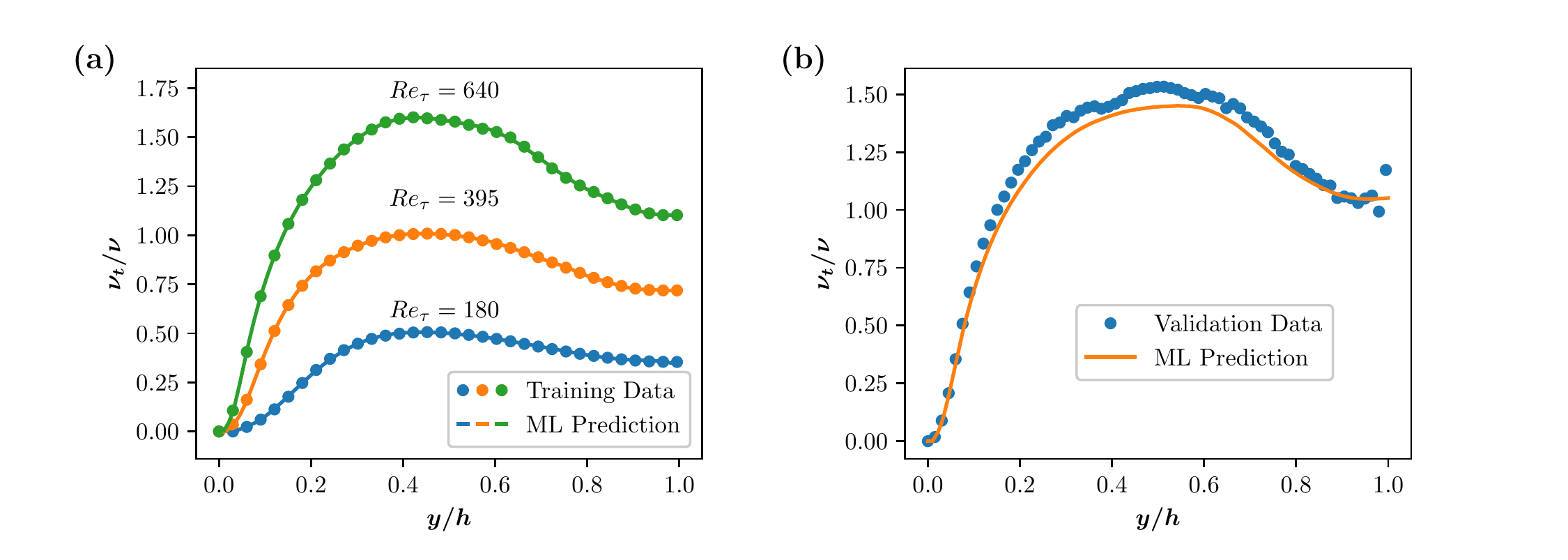}%
    \caption{\label{Fig:Priorall_nt} The eddy-viscosity fitted by the NN at the end of the training phase.
(a): Performance of the training set (b): Performance of the validation set}
\end{figure}

\begin{table}[h!]
    \caption{ Hyperparameters of a successful training of the NN \label{Table:3}}
    \centering
    \begin{tabular*}{\textwidth}
        {@{\hspace{6em}}c @{\extracolsep{\fill}}c @{\hspace{6em}}}
    \\[-11pt]
    \toprule 
    \textbf{Hyperparameters}     & \textbf{Recommend Values} \\[1pt] \hline \\[-9pt]
    Number of Hidden Layers       & 3                         \\
    Number of Nodes per Layer     & 24                        \\
    Activation Function           & tanh                      \\
    Optimizer                     & Adam Optimizer            \\
    Learning Rate                 & 0.002                     \\
    L1 Regularization Coefficient & 0.025                     \\
    Epoch Numbers                 & 1500000                   \\
    \bottomrule  
    \end{tabular*}
\end{table}

In Fig.~\ref{Fig:Errors_nt}, the quality of the training dataset is observed to be better than the validation dataset 
in terms of the number of samples for averaging and the smoothness of the high-order statistics. Consequently,
the magnitude of MSE of the validation data is larger than that of the training data. However, at the end of
the training, the error of the validation data converges to the lowest level, indicating that the ML model
is not over-fitted in the training process.

The \textit{a priori} result of the eddy-viscosity is presented in Fig.~\ref{Fig:Priorall_nt}. It can be seen
that the eddy-viscosity is perfectly predicted in the training dataset, and a satisfactory result is achieved
in the validation dataset as well.

During the training phase, the overfitting problem is avoided by using a set of noisy data to monitor and control
the training process, leading to a satisfactory \textit{a priori} prediction of both
the training and the validation sets. The details of the hyperparameters in the NN are listed in Table~\ref{Table:3}.

\subsection{A Posteriori Results in Test Cases}
After the training stage, the weights of each layer in the NN are frozen and the ML model is ready
for an online prediction. In the prediction phase, the NN model saved in the frozen Tensorflow graph
is loaded to the CFD solver to predict the eddy-viscosity at each iteration step based on the input features
composed with the runtime flow field.

\subsubsection{Planar Turbulent Channel Flows}

\begin{figure}[!ht]
    \centering
    \includegraphics[width=0.77\textwidth,trim=60 198 110 182,clip]{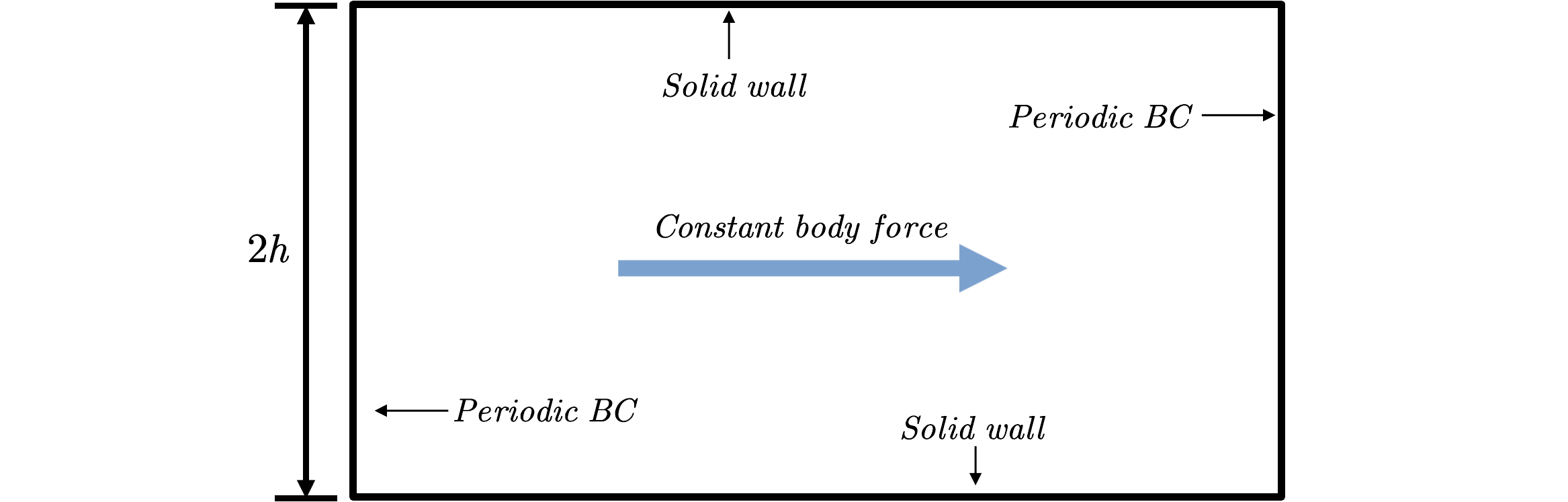}%
    \caption{\label{Fig:Geo_Channel}The geometry and boundary conditions for channel flow simulation}
\end{figure}

\begin{figure}[!h]
    \centering
    \includegraphics[width=0.70\textwidth,trim=0 5 0 35,clip]{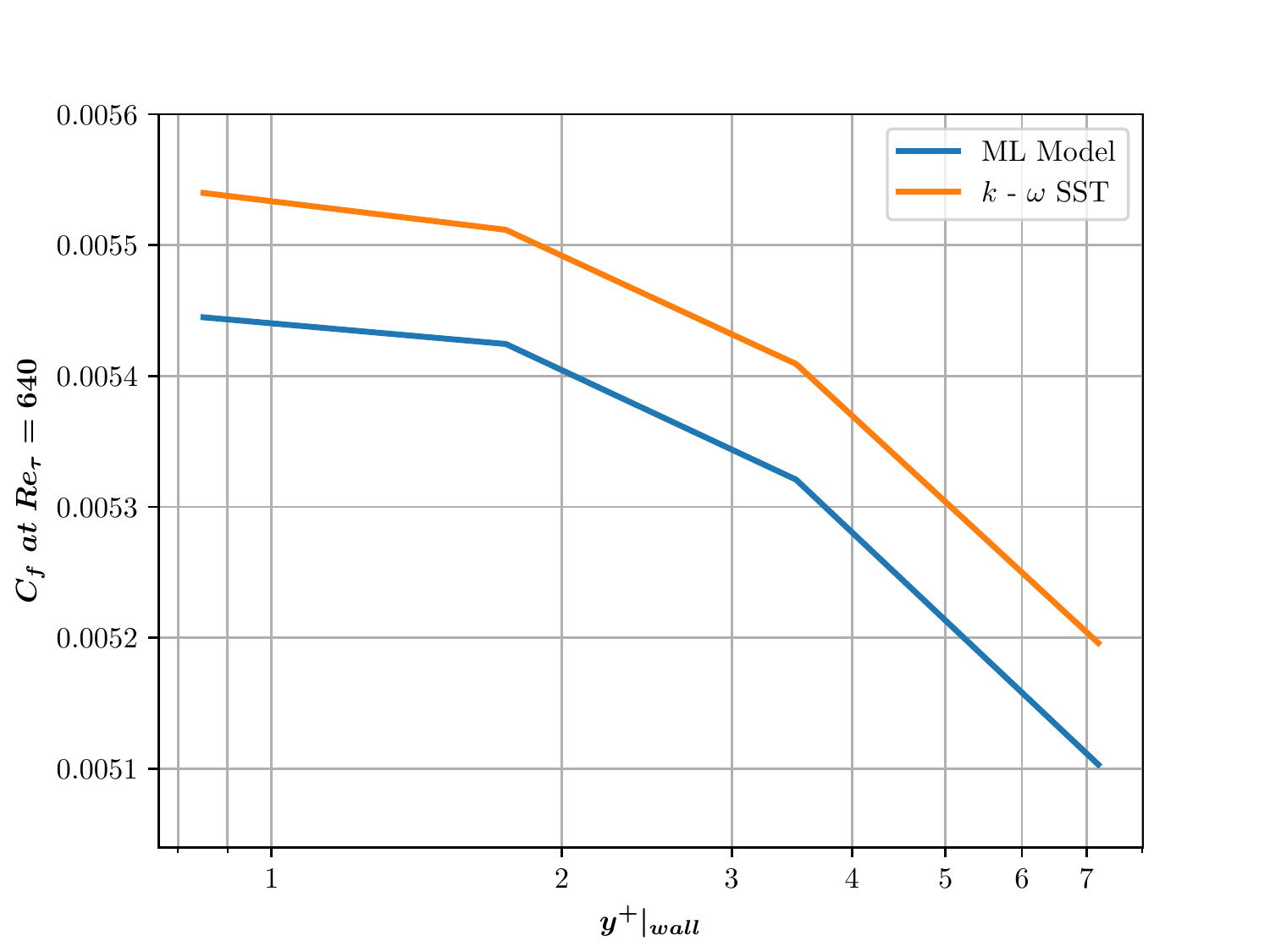}%
    \caption{\label{Fig:Mesh_Channel}Grid convergence for channel flow at $Re_\tau$= 640}
\end{figure}

\begin{figure}[!h]
    \centering
    \includegraphics[width=0.85\textwidth,trim=0 5 0 25,clip]{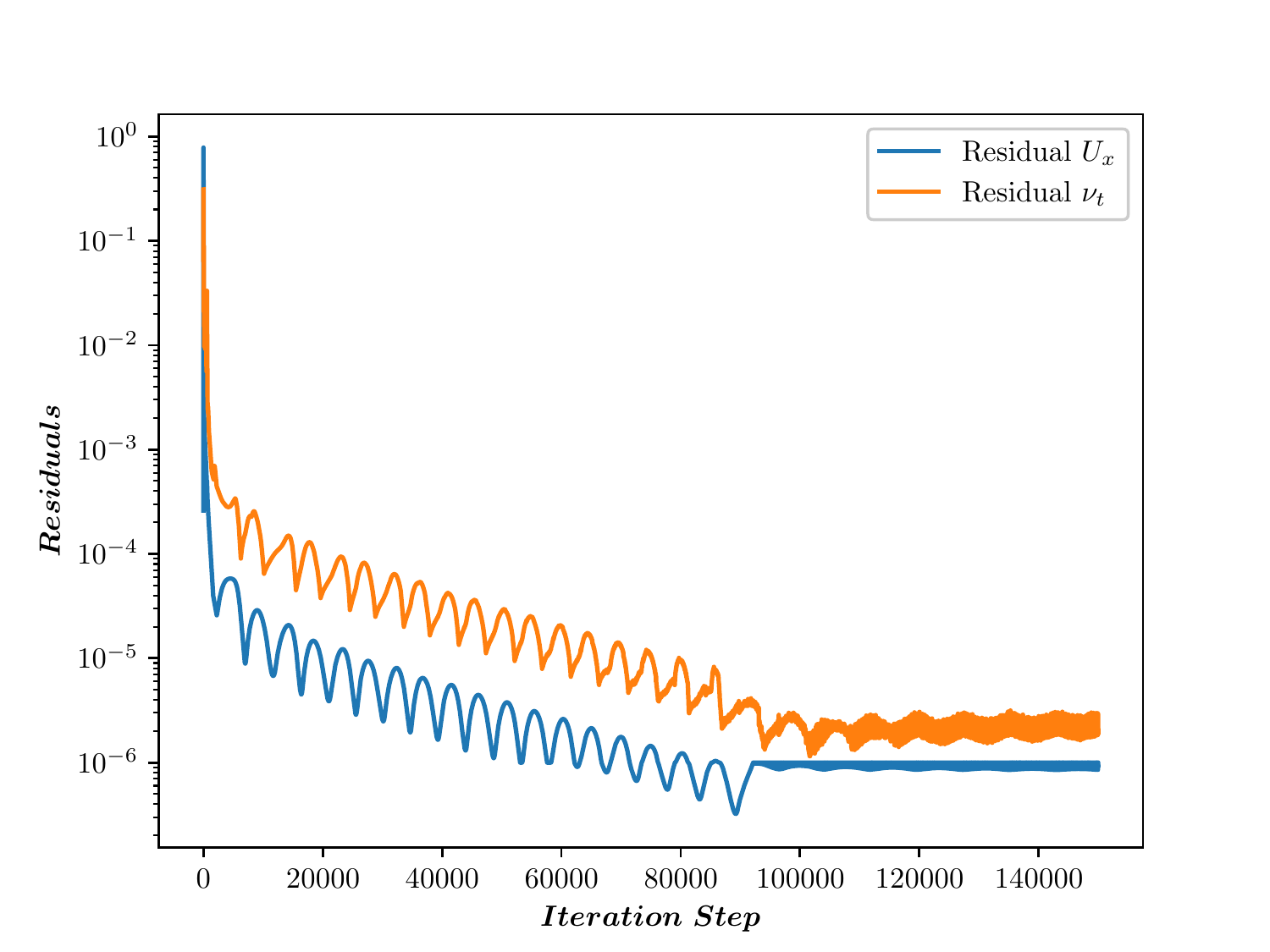}%
    \caption{\label{Fig:Residual_nt}Evolution of residuals in the simulation of a channel flow.}
\end{figure}

The developed ML model is tested in turbulent channel flows, which is depicted
in Fig.~\ref{Fig:Geo_Channel}.
A grid convergence study of the channel flow with both $k$--$\omega$ SST model and the ML model is shown
in Fig.~\ref{Fig:Mesh_Channel}, from which we can observe converged friction coefficients when $y^+$
in the first layer is less than 2. Consequently, the meshes listed in Table~\ref{Table:4}
have been adopted in the \textit{a posteriori} cases with the first point off the wall
satisfying $y^+|_{wall}\approx \,\,1$ and the maximal mesh resolution $\varDelta y_{\max}^{+}<\,\,4$.
The initial field for each test is the converged solution from $k$--$\omega$ SST model.
The residuals of velocity and eddy-viscosity are monitored in Fig.~\ref{Fig:Residual_nt}.

\begin{table}[!ht]
    \caption{Mesh details of RANS simulation \label{Table:4}}
    \centering
    \begin{tabular*}{\textwidth}
        {@{\hspace{8em}}c @{\extracolsep{\fill}}c @{\hspace{8em}}}
    \\[-11pt]
    \toprule 
    \textbf{\begin{tabular}[c]{@{}c@{}}Reynolds Number\\ ($Re_\tau$)\end{tabular}} &
    \textbf{\begin{tabular}[c]{@{}c@{}}Cell Number\\ (wall normal)\end{tabular}} \\[1pt] \hline \\[-9pt]
    150 \textless{} $Re_\tau$ \textless{} 300                       & 168               \\
    300 \textless{} $Re_\tau$ \textless{}  500                      & 372               \\
    $Re_\tau$ \textgreater{}  500                                   & 558               \\
    \bottomrule                                                            
    \end{tabular*}
\end{table}

\begin{figure}[!h]
\centering
\includegraphics[width=0.9\textwidth,trim=0 17 0 22,clip]{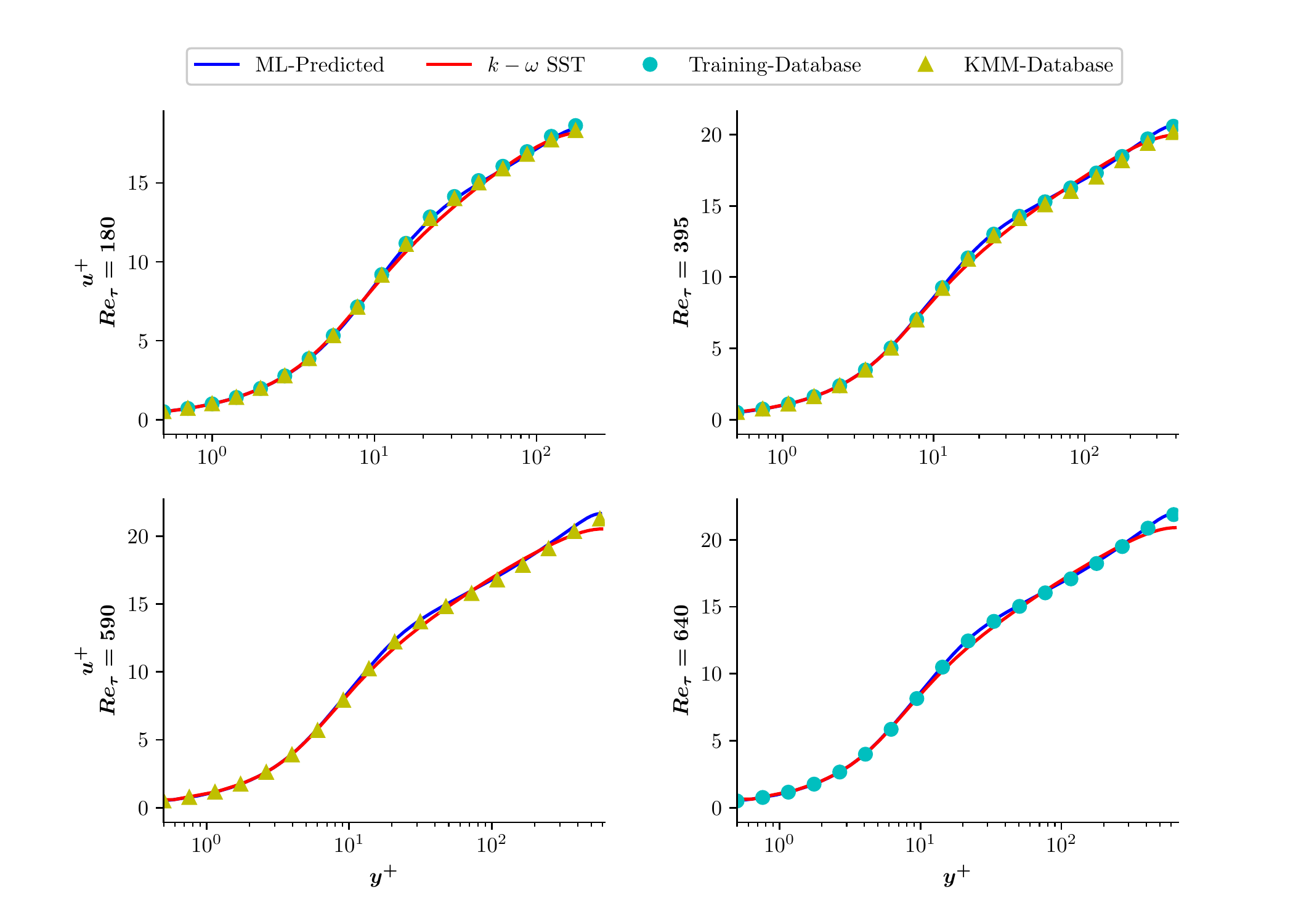}%
\caption{\label{Fig:Posterior_u_nt}Comparison of velocity profile.}
\end{figure}

\begin{figure}[!h]
\centering
\includegraphics[width=0.9\textwidth,trim=0 17 0 22,clip]{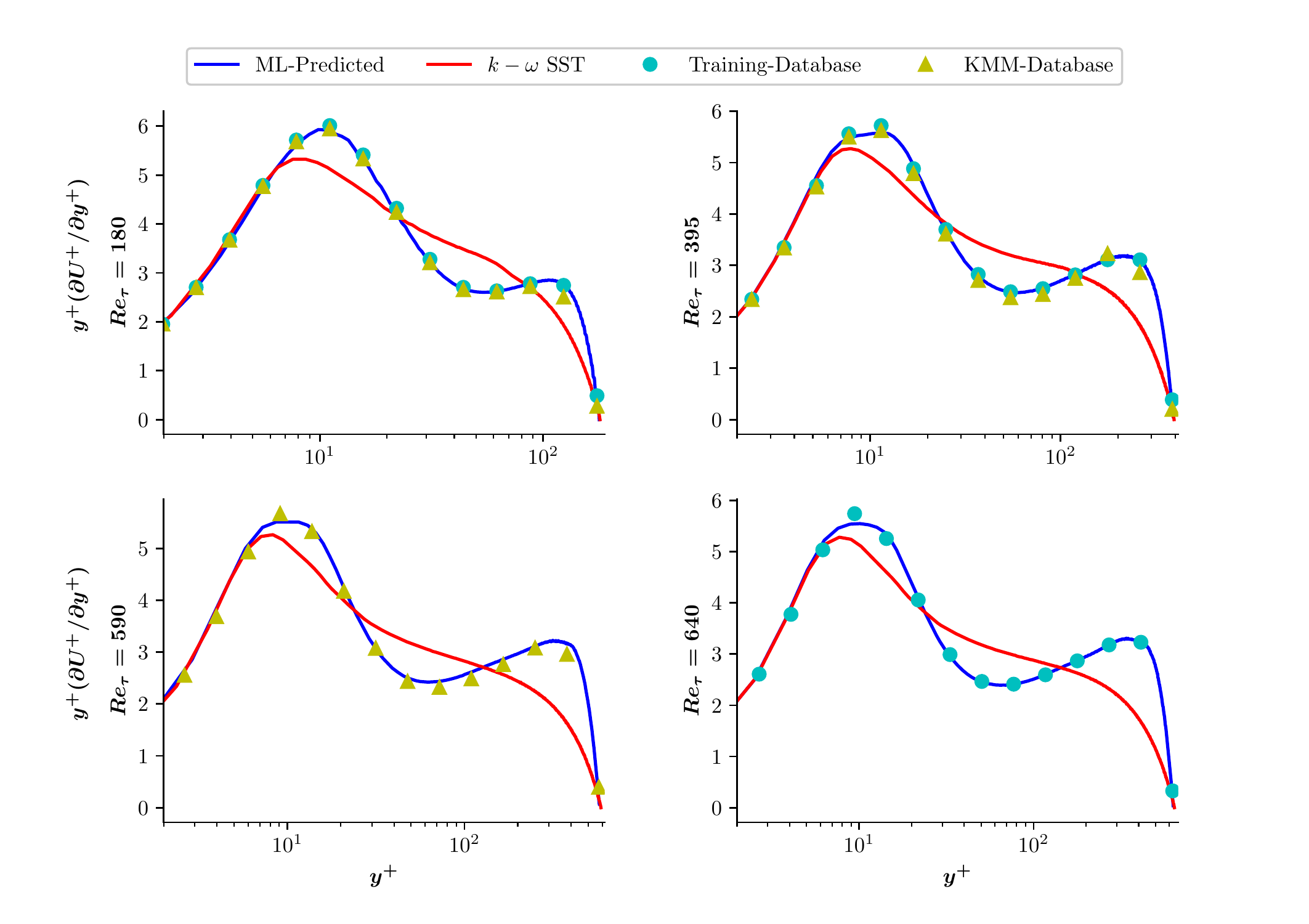}%
\caption{\label{Fig:Posterior_ududy_nt}Comparison of $y^+(\partial{U^+}/\partial{y^+})$ distributions.}
\end{figure}

\begin{figure}[!h]
\centering
\includegraphics[width=0.9\textwidth,trim=0 17 0 22,clip]{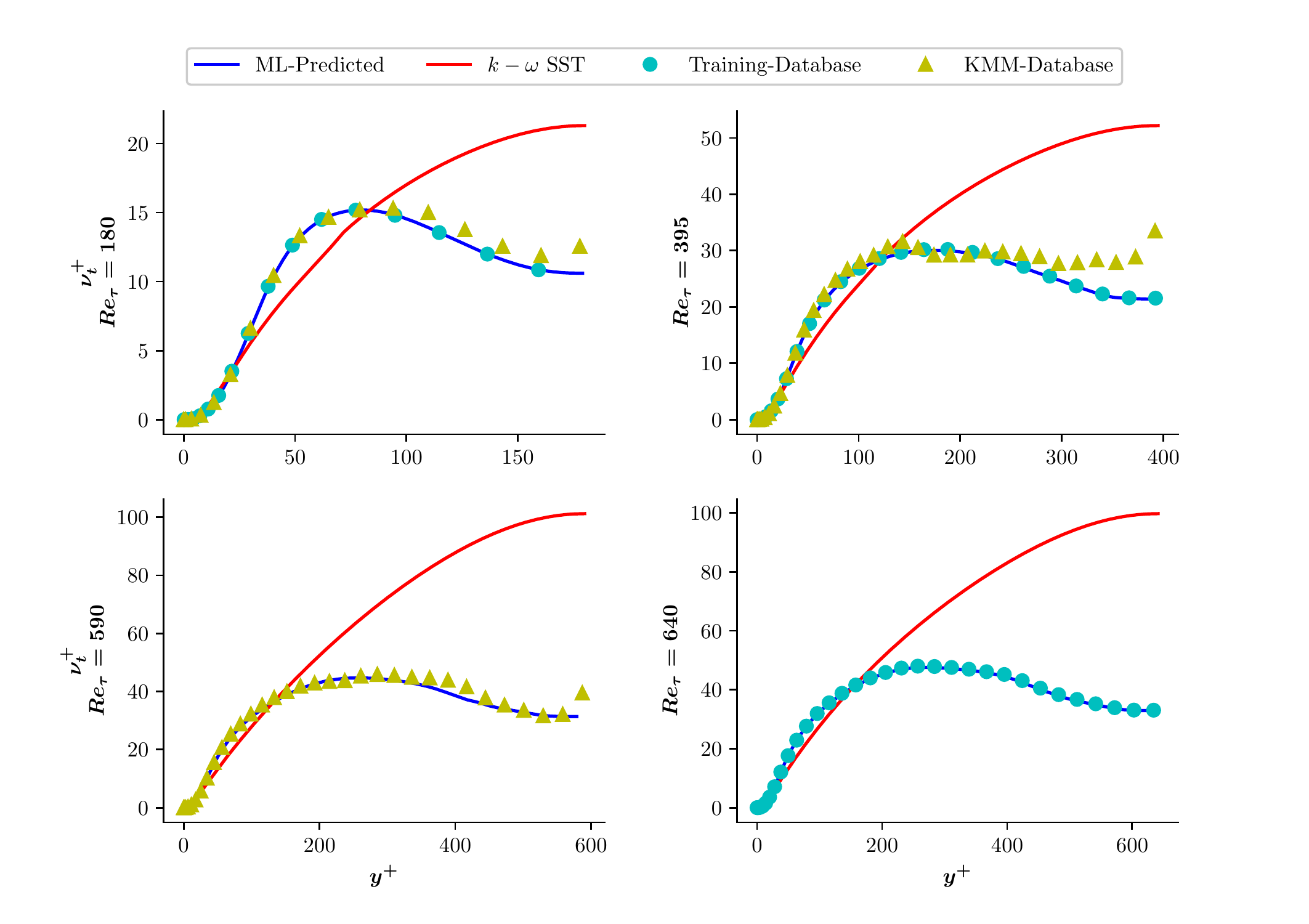}%
\caption{\label{Fig:Posterior_nut_nt_sec}Comparison of eddy-viscosity distributions along wall normal direction.}
\end{figure}

The profiles of mean velocity and normalized mean velocity gradient, $y^+(\partial U^+ /\partial y^+ )$,
from the \textit{a posteriori} simulations are shown in Fig.~\ref{Fig:Posterior_u_nt}
and Fig.~\ref{Fig:Posterior_ududy_nt}, respectively. As expected, the mean velocity profiles are predicted
accurately by the ML integrated turbulence model, with no noticeable difference with the DNS data. From the
detailed comparison of velocity gradient profiles, we can further confirm that the traditional $k$--$\omega$
SST model failed to capture the peak and inflection points of the velocity gradient~\cite{Sylvaindevelopment2016}, 
and the proposed ML model presents a superior performance in predicting high-order statistics.

The eddy-viscosity profiles are compared in Fig.~\ref{Fig:Posterior_nut_nt_sec}, from which we can see
that the $k$--$\omega$ SST model shows an over-prediction of the eddy-viscosity~\cite{Sylvaindevelopment2016},
leading to errors in the predicted mean velocity. Our ML model improves the results significantly, and the
profiles of eddy-viscosity for all cases agree well with the DNS data. The wiggles on the eddy-viscosity
profiles of the KMM database might be due to the inadequacy of samples for averaging. (A contrast to major traditional
models can be found in~\ref{Ap:B})

An analysis for quantitative errors is also performed under the definition of the root-mean-square
(\textsl{r.m.s.}) error of streamwise velocity in channel flows, given by Eq.~(\ref{eq:18}), and the  \textsl{r.m.s.}
errors are compared in Table~\ref{Table:5}

\begin{equation}
e_c=\dfrac{\sqrt{\dfrac{1}{R e_{\tau}} \displaystyle \int_{0}^{R e_{\tau}}\left(U_{d n s}^{+}-U_{r a n s}^{+}
\right)^{2} d y^{+}}}{\dfrac{1}{R e_{\tau}} \displaystyle \int_{0}^{R e_{\tau}} U_{d n s}^{+} d y^{+}} \label{eq:18}
\end{equation}

\begin{table}[!h]
    \caption{  \textsl{r.m.s.} errors compared with DNS at same $Re_\tau$ \label{Table:5}}
    \centering
    \begin{tabular}{ccccccc}
    \toprule 
    \textbf{Turbulence} & \multicolumn{3}{c}{\textbf{DNS~Abe~\textit{et al.}~\cite{RN62}}}  & \multicolumn{3}{c}{\textbf{KMM database~\cite{RN51}}}    \\ 
    \textbf{model}    & $Re_\tau=180$  & $Re_\tau=395$ & $Re_\tau=640$ & $Re_\tau=180$ & $Re_\tau=395$ & $Re_\tau=590$  \\ [2pt]\hline \\[-10pt]
    $k$--$\omega$ SST & 1.27\%         & 1.74\%        & 2.36\%        & 1.26\%        & 1.64\%        & 2.13\%         \\
    ML model          & 0.12\%         & 0.08\%        & 0.08\%        & 0.29\%        & 0.20\%        & 0.26\%         \\       
    \bottomrule  
    \end{tabular}
\end{table}

As Table~\ref{Table:5} shows, the ML model outperforms the conventional $k$--$\omega$ SST model with the
error being an order of magnitude lower, indicating that a promising reproducibility of the training cases
can be achieved in the present framework.

\begin{figure}[!h]
\centering
\includegraphics[width=1\textwidth,trim=0 0 0 0,clip]{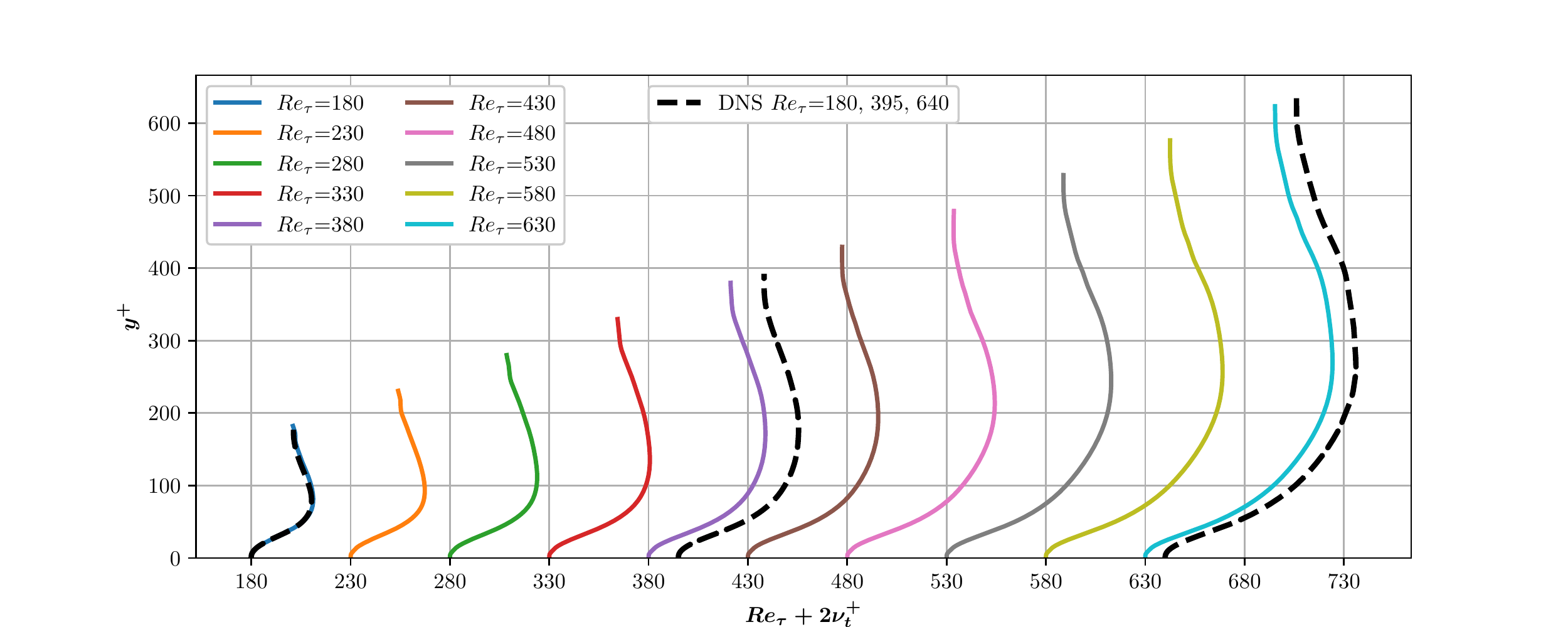}%
\caption{\label{Fig:Robust_Fine_nut_nt} Distributions of eddy-viscosity in channel flows at equally spaced
Reynolds numbers.}
\end{figure}

\begin{figure}[!h]
\centering
\includegraphics[width=0.8\textwidth,trim=0 5 0 32,clip]{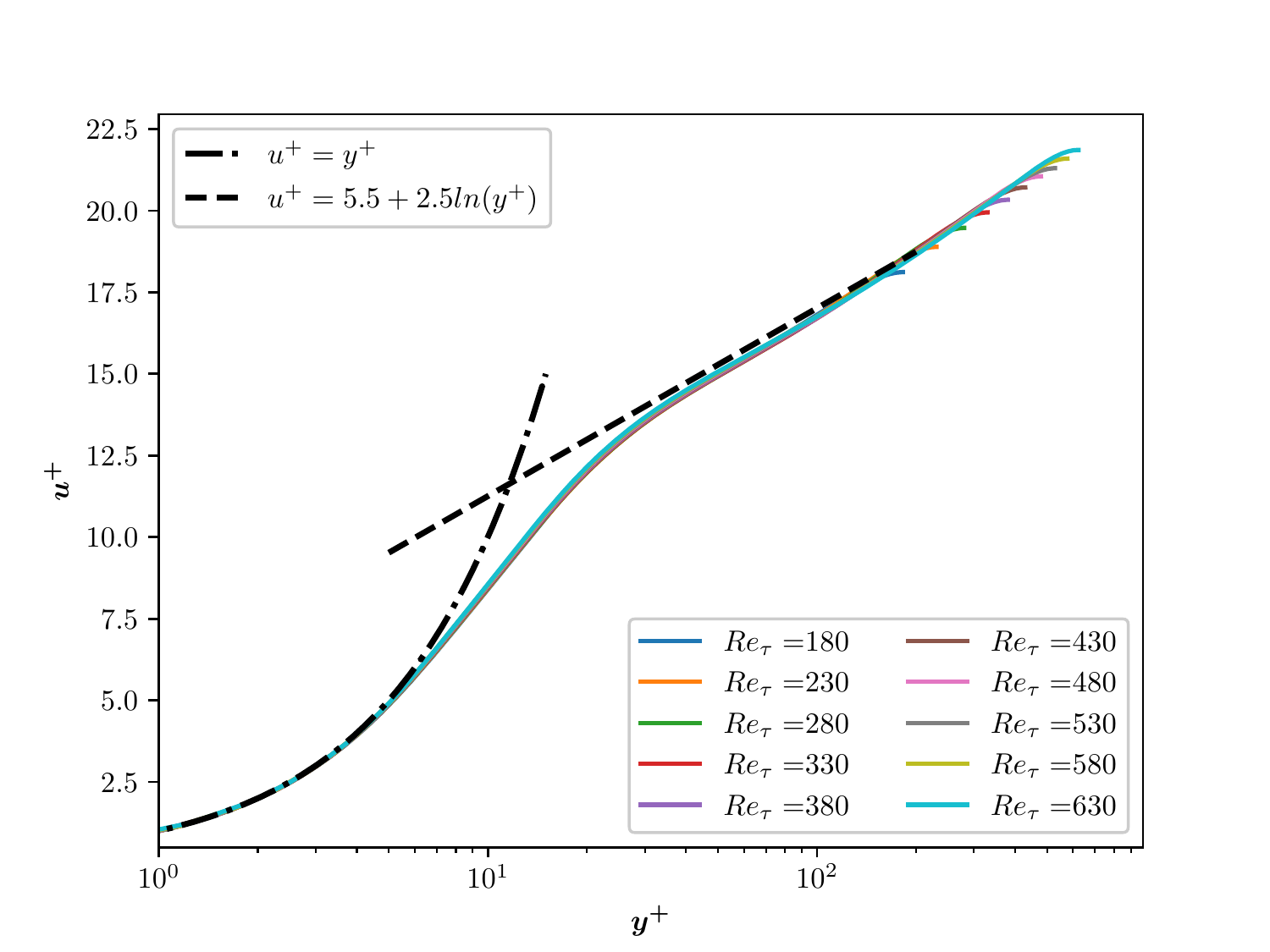}%
\caption{\label{Fig:Robust_Fine_nt} Mean velocity profiles in channel flows at equally spaced
Reynolds numbers.}
\end{figure}

\begin{figure}[!h]
\centering
\includegraphics[width=0.8\textwidth,trim=11 12 0 10,clip]{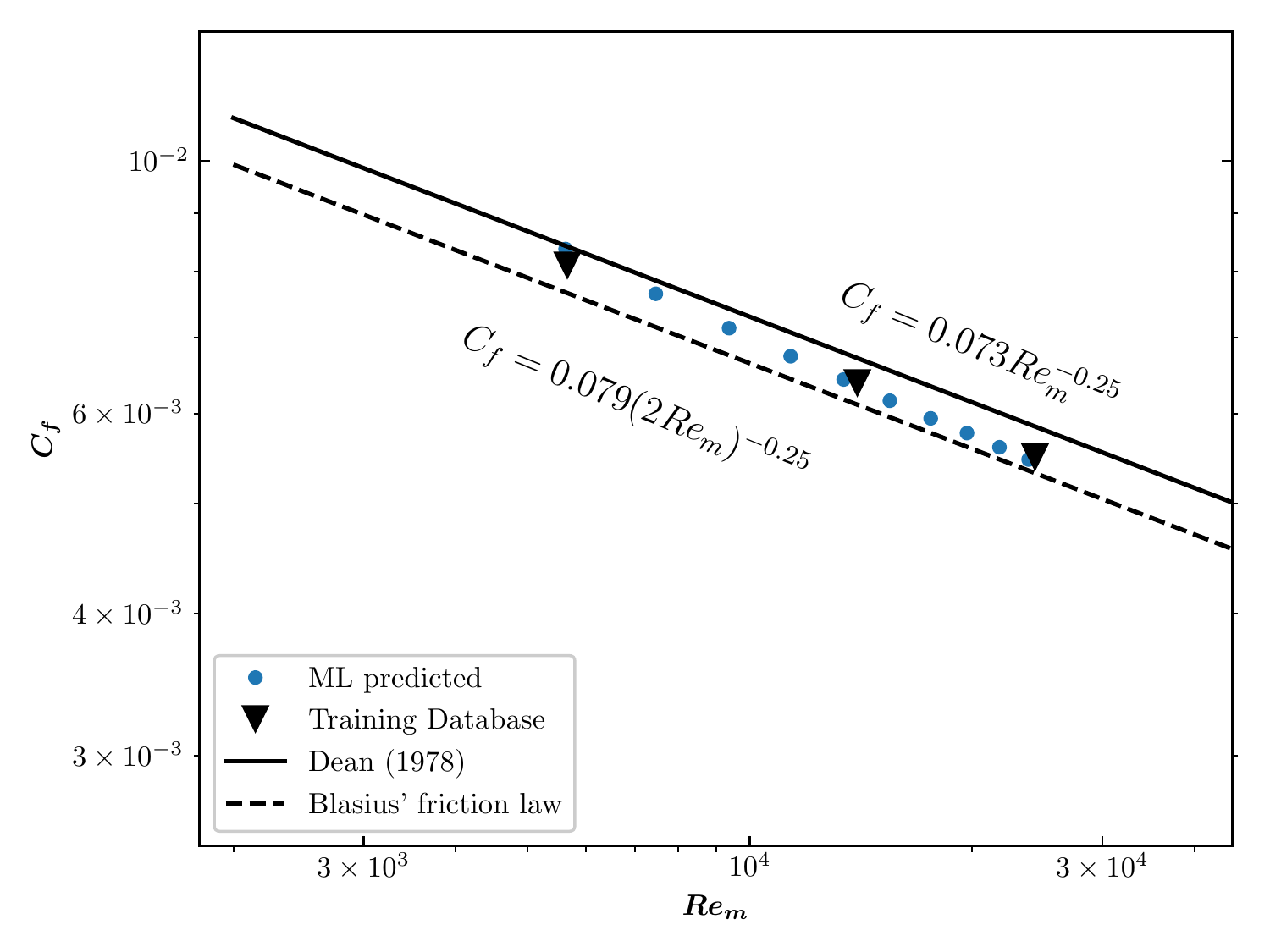}%
\caption{\label{Fig:Channeltau_nt} Comparison of friction coefficients of ML model’s prediction, DNS and
analytical formula, where $Re_m$ is the Reynolds number based on the bulk velocity and height of the 
channel.}
\end{figure}

To further evaluate the robustness and interpolation capability of the developed ML model,
we also test the ML-RANS model in a channel flow at Reynolds number changing continuously from $Re_\tau$ = 180
to $Re_\tau$ = 630. The profiles of eddy-viscosity and mean velocity are presented in Fig.~\ref{Fig:Robust_Fine_nut_nt}
and Fig.~\ref{Fig:Robust_Fine_nt} respectively, and the skin friction coefficients are presented
in Fig.~\ref{Fig:Channeltau_nt}. It can be observed that the linear law and the log law of the velocity
profile are well preserved in all the tests, and the friction coefficients agree well with the law of
Dean~\cite{RN103} and the Blasius friction law. From the results of this series of simulations, we can see a
consistent changing of profiles with the Reynolds number, demonstrating favorable robustness and interpolation
capability of the ML model within the range of training data.

From the \textit{a posteriori} results, we have shown that the developed ML-RANS framework not only obtains an iteratively converged simulation but also reproduces the mean flow field from the training
cases. The ML-RANS framework can further perform a satisfactory interpolation capability at Reynolds
numbers within the range of training data. Therefore, good reproducibility of the training cases and a favorable interpolation
capability can be achieved in the present ML-RANS framework.

\subsubsection{Flow over Periodic Hills}

\begin{figure}[!ht]
    \centering
    \includegraphics[width=0.8\textwidth,trim=30 200 100 135,clip]{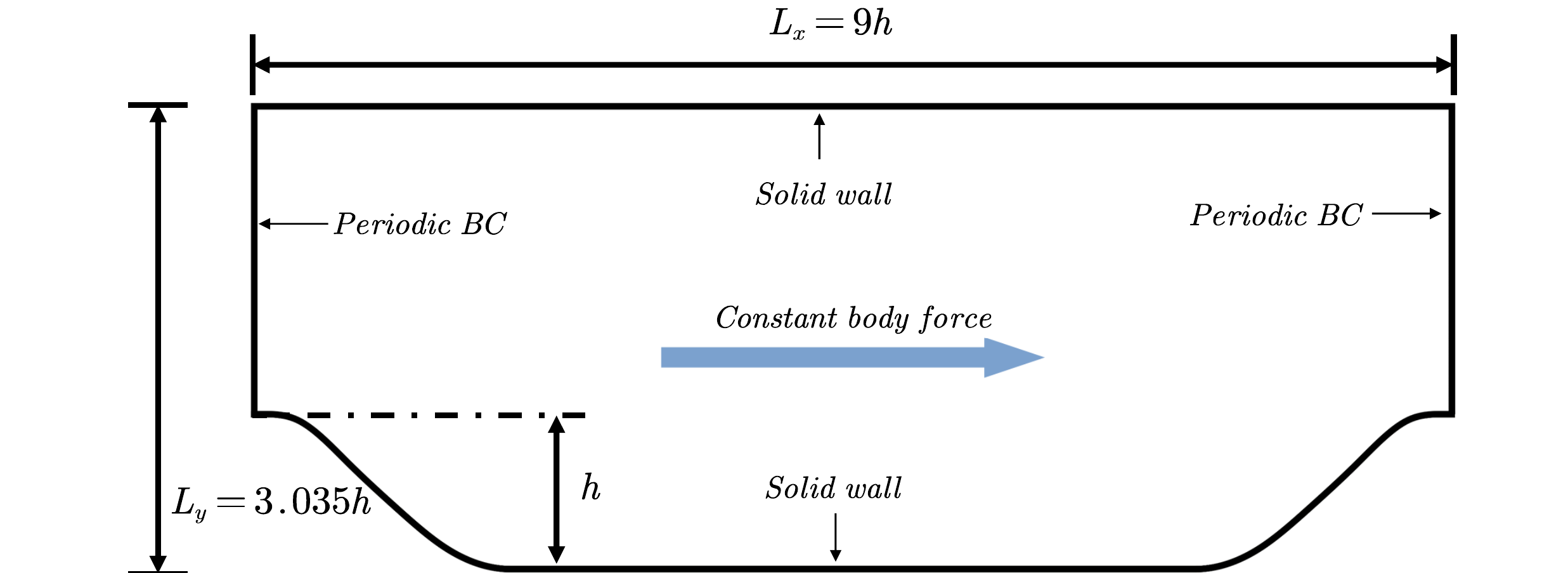}%
    \caption{\label{Fig:Geo_Phill} The geometry and boundary condition for flow over periodic hills.}
\end{figure}

\begin{figure}[!ht]
    \centering
    \includegraphics[width=0.8\textwidth,trim=15 165 5 183,clip]{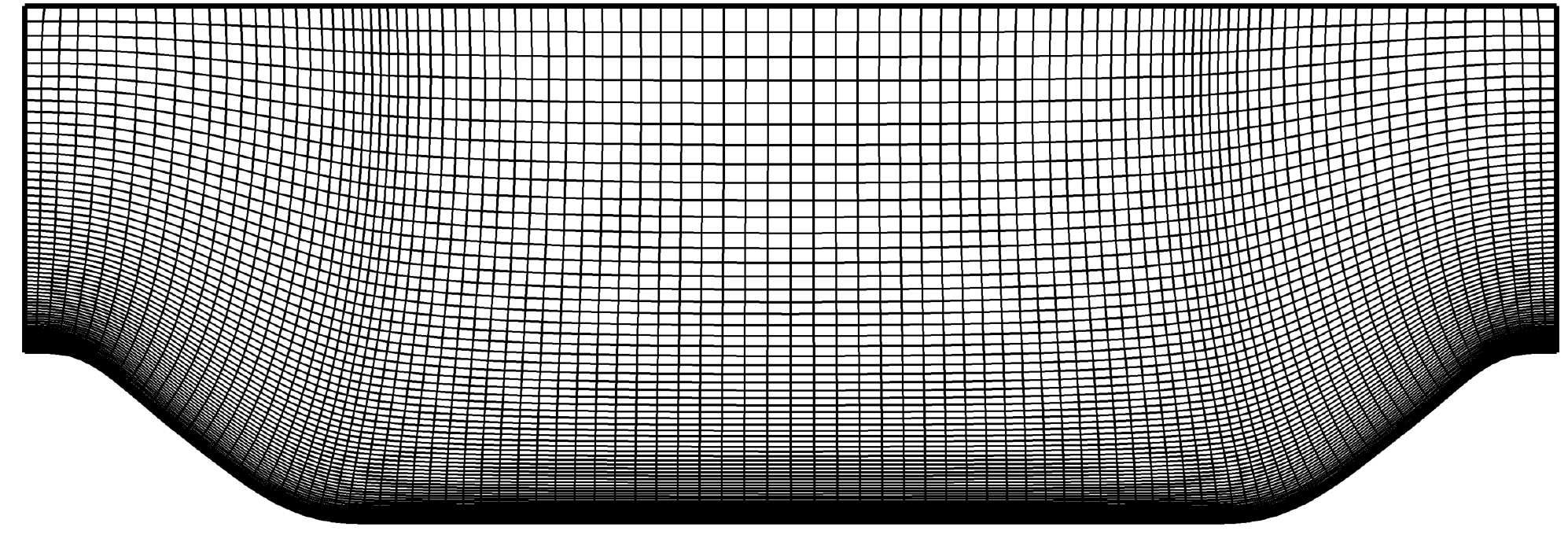}%
    \caption{\label{Fig:Grid_Phill} The body-fitted grid used in the simulation, each 4 grid line is shown in the total dimension of $393\times257$.}
\end{figure}

\begin{figure}[!ht]
    \centering
    \includegraphics[width=1\textwidth,trim=100 70 100 60,clip]{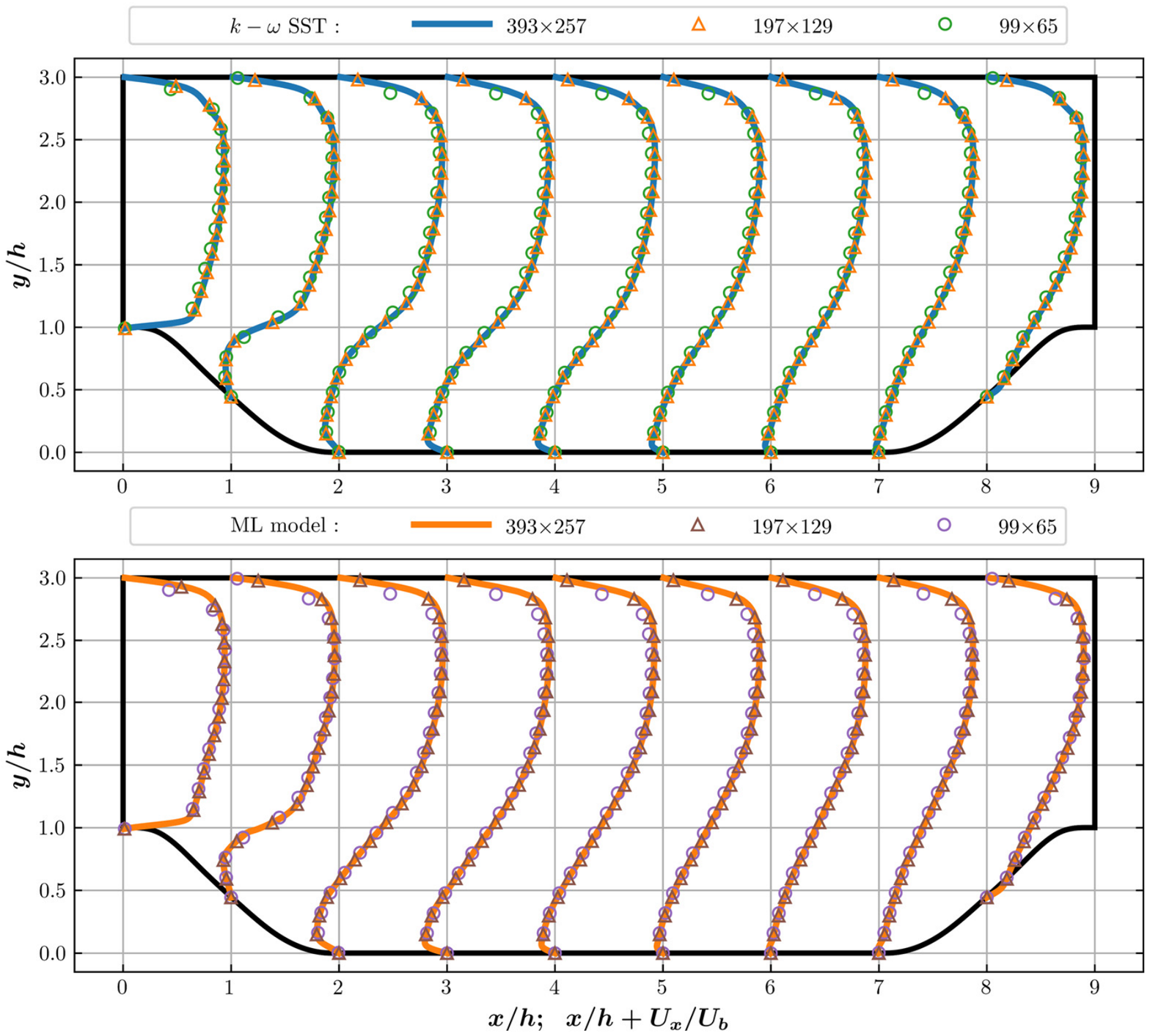}%
    \caption{\label{Fig:Mesh_Phill} Steamwise velocity profile in flow over periodic hills with different
grid resolution. }
\end{figure}

The ML-RANS model trained only with data from channel flows will now be applied to a more complex case 
that involves separation, since its input and output variables are all locally-defined and independent by flow cases. 
Therefore, flow over periodic hills at $Re_H$ = 1400 is conducted to further examine the performance of the ML model.
The geometry and boundary conditions are shown in Fig.~\ref{Fig:Geo_Phill}, and a $393\times257$ body-fitted grid with $y_1^+\approx0.12$ on the bottom wall (see Fig.~\ref{Fig:Grid_Phill}) is adopted to discretise the domain. The grid independence study shown in  Fig.~\ref{Fig:Mesh_Phill} indicates the grid convergence is achieved for RANS simulations with both $k$--$\omega$ and ML models.

\begin{figure}[!ht]
\centering
\includegraphics[width=1.05\textwidth,trim=40 70 70 60,clip]{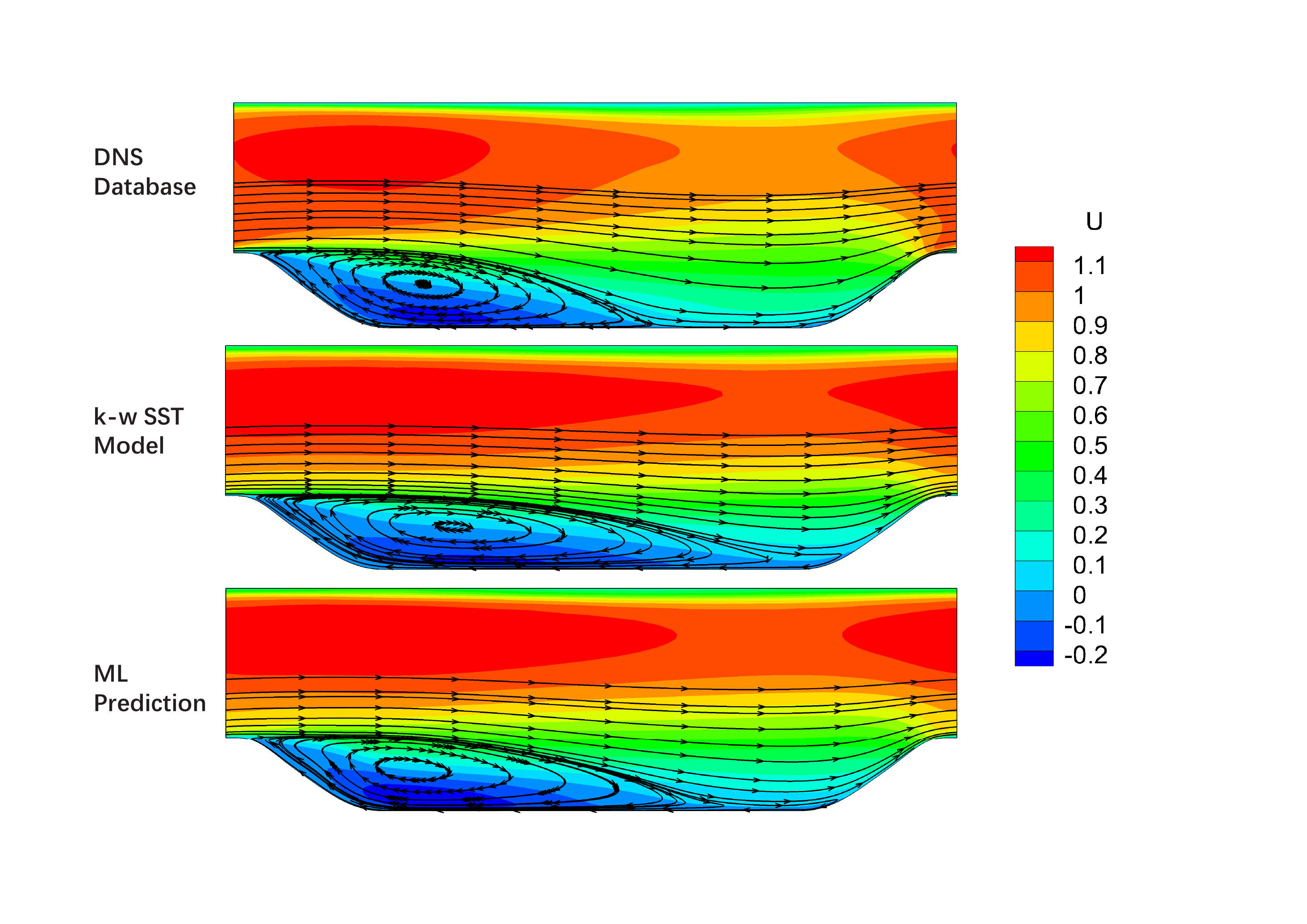}%
\caption{\label{Fig:Velo} Mean velocity field in flow over periodic hills.}
\end{figure}

\begin{figure}[!ht]
\centering
\includegraphics[width=0.96\textwidth,trim=0 25 0 0,clip]{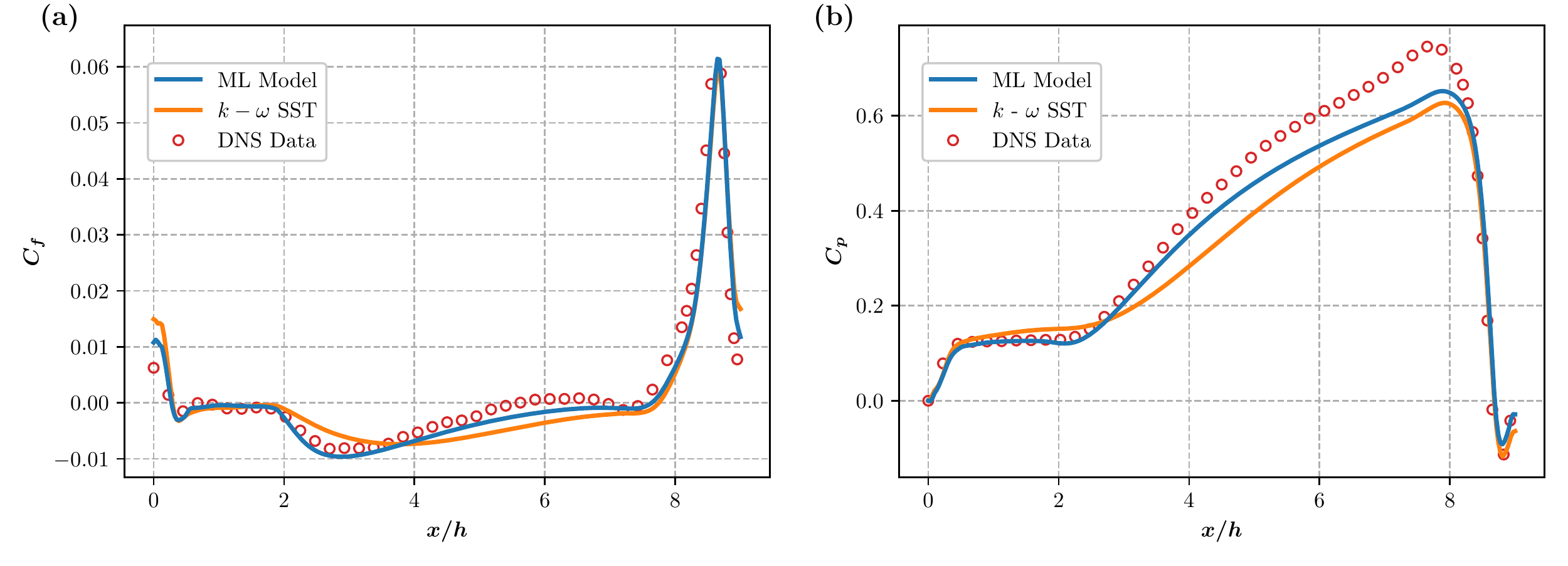}%
\caption{\label{Fig:cfcp} Distribution of friction coefficient (a) and pressure coefficient (b) on the bottom wall
of the periodic hill. }
\end{figure}

For this particular problem, the ML model shows a clear improvement against the traditional model, even though it has been trained using data from planar channel flows. The comparison of the mean velocity fields from DNS, $k$--$\omega$ SST model,
and the ML model is presented in Fig.~\ref{Fig:Velo}. The ML model presents a better result than the
original $k$--$\omega$ SST model in terms of the size of the separation bubble and the location of the
reattachment point. A comparison of mean skin friction and pressure coefficients on the bottom wall also indicate an improved performance of the ML model, as shown in Fig.~\ref{Fig:cfcp}. 

The uncertainty in the separated case might come from the isotropic nature of the eddy-viscosity failing to
predict the condition where the principal axis of the Reynolds stress tensor deviates from those of the strain rate tensor.
However, these results also show great potential for the ML model to achieve a favorable prediction in complex
flows, given enough training from simple cases. 

 As anticipated, the ML model developed in the present study succeeds in the simulation of the same type of
flow as the training case and it also shows good potential in predicting a more complicated flow. This result
is very encouraging for the further development and application of the ML model.

\subsection{Computation time}
Due to different hardware and computational configuration, the absolute running time of a solver may differ, but the comparison of the computation time to a traditional model within the same configuration is still informative. The wall clock times of the ML model and $k$--$\omega$ SST model for one iteration in a channel flow are compared in Table 6. The mesh count was artificially increased by adding cells in the streamwise and wall-normal direction to increase the wall clock time and make it more significant for a general application. The tests are conducted on a single core of \verb|Intel(R) Xeon(R) CPU E5-2630 v4 2.20GHz|.

\begin{table}[!ht]
    \caption{Wall clock time cost for a single iteration}
    \centering
    \begin{tabular}
        {@{\hspace{3em}}c @{\extracolsep{\fill}}ccc @{\hspace{3em}}}
    \\[-11pt]
    \toprule 
    \textbf{Mesh} &\textbf{Cell number} & \textbf{$k$--$\omega$ SST} & \textbf{ML Model}   \\
    [1pt] \hline \\[-9pt]
    40 $\times$ 169  &6552           & 0.02s  & 0.03s \\
    80 $\times$ 373  &29388          & 0.07s  & 0.15s \\
    130 $\times$ 559 &71982          & 0.17s  & 0.35s \\
    \bottomrule  
    \end{tabular}
\end{table}

It can be observed that the computation time of the ML model is roughly twice of the $k$--$\omega$ SST model, since the transport equations of the $k$--$\omega$ SST model are also solved in the ML model. It also indicates that the forward inference of NN consumes approximately the same computation time as solving two PDEs, which is quite acceptable for a RANS simulation.

\section{Conclusions - Perspectives\label{Sec:Conclusion}}
In this study, an iterative ML-RANS framework is developed aiming at a built-in reproducibility of the
training cases. First, a conventional RANS model (the $k$--$\omega$ SST model is used in the present study) is incorporated
to process the training data so that the calculation of input features in both the training phase
and the predicting phase are conducted in the same manner. This also brings an empirical estimation for
turbulence quantities into the system, making it possible to compose more independent variables for the ML
regression system. To overcome the ill-conditioning problem of RANS equations, the Reynolds stress tensor is decomposed into the linear part and the residual nonlinear part, and the linear part is treated implicitly and incorporated into the diffusion terms of the RANS equations. Furthermore, the existing constitutive hypothesis is extended with more independent variables to ensure the
sufficiency of the input feature of the ML model so that a single-valued regression system suitable for the ML algorithm can be established. 

A program interface is developed to couple the machine learning library to a CFD solver, and an iterative convergence is achieved in the \textit{a posteriori} simulations. The developed ML-RANS framework is trained under a cross-case strategy, and DNS data from turbulent channel flows at $Re_\tau$ = 180, 395, and 640 are merged together and fed to the ML model. 
This framework is tested against channel flow with equally-spaced Reynolds numbers that are within the initial training range (i.e. 180 $\leq Re_\tau \leq$ 640) as well as the flow over periodic hills. The results show that the ML model not only achieves a good reproducibility of the training cases but also performs a favorable interpolation capability within the range of the training data. For the test case involving flow over periodic hills, the ML model outperforms the $k$--$\omega$ SST model,
even though the model has not been provided with any prior information about the flow. This predictive
capability indicates a promising potential of the current ML-RANS framework. Moreover, the computational cost of the NN is at the same level of solving two PDEs.

Although the $k$--$\omega$ SST model has been incorporated in the current study, the framework is compatible with other turbulence models. Further developments are still on-going, including the way of obtaining continuous decomposed fields for closure terms in complex flows, and a method to detect multi-valued phenomena in high-dimensional data. 

\appendix
\section{The Entire Tensor Bases for Algebraic Stress Model  \label{Ap:A}}
The full expression in the explicit constitutive equation can be found as follows:

\begin{equation}
	\boldsymbol{b^*} =\sum_{m=1}^{10}{G^{\left( m \right)}}\left( \lambda _1,...,\lambda _5 \right) \boldsymbol{T}^{\left( m \right)}  
\end{equation}

where the definition of $\boldsymbol{T}^{(m)}$ and $\lambda$ are defined as: 

$$\boldsymbol{T}^{\left( m \right)}=\left\{ \begin{array}{c}	\begin{array}{l}	\boldsymbol{T}^{\left( 1 \right)}=\boldsymbol{{S}^*}\\	\\	\boldsymbol{T}^{\left( 2 \right)}=\boldsymbol{{S}^*{\varOmega}^*}-\boldsymbol{{\varOmega}^*{S}^*}\\	\\	\boldsymbol{T}^{\left( 3 \right)}=\boldsymbol{{S}^*}^2-{1}/{3}\boldsymbol{E}\cdot tr\left( \boldsymbol{{S}^*}^2 \right)\\	\\	\boldsymbol{T}^{\left( 4 \right)}=\boldsymbol{{\varOmega}^*}^2-{1}/{3}\boldsymbol{E}\cdot tr\left( \boldsymbol{{\varOmega}^*}^2 \right)\\	\\	\boldsymbol{T}^{\left( 5 \right)}=\boldsymbol{{\varOmega}^*{S}^*}^2-\boldsymbol{{S}^*}^2\boldsymbol{{\varOmega}^*}\\\end{array}\\\end{array}\,\,     \right. \begin{array}{l}	\boldsymbol{T}^{\left( 6 \right)}=\boldsymbol{{\varOmega}^*{S}^*}^2+\boldsymbol{{S}^*}^2\boldsymbol{{\varOmega}^*}-{2}/{3}\boldsymbol{E}\cdot tr\left( \boldsymbol{{S}^*{\varOmega}^*}^2 \right)\\	\\	\boldsymbol{T}^{\left( 7 \right)}=\boldsymbol{{\varOmega}^*{S}^*{\varOmega}^*}^2-\boldsymbol{{\varOmega}^*}^2\boldsymbol{{S}^*{\varOmega}^*}\\	\\	\boldsymbol{T}^{\left( 8 \right)}=\boldsymbol{{S}^*{\varOmega}^*{S}^*}^2-\boldsymbol{{S}^*}^2\boldsymbol{{\varOmega}^*{S}^*}\\	\\	\boldsymbol{T}^{\left( 9 \right)}=\boldsymbol{{\varOmega}^*}^2\boldsymbol{{S}^*}^2+\boldsymbol{{S}^*}^2\boldsymbol{{\varOmega}^*}^2-{2}/{3}\boldsymbol{E}\cdot tr\left( \boldsymbol{{S}^*}^2\boldsymbol{{\varOmega}^*}^2 \right)\\	\\	\boldsymbol{T}^{\left( 10 \right)}=\boldsymbol{{\varOmega}^*{S}^*}^2\boldsymbol{{\varOmega}^*}^2-\boldsymbol{{\varOmega}^*}^2\boldsymbol{{S}^*}^2\boldsymbol{{\varOmega}^*}\\ \end{array},$$
$$\lambda _1=tr\left( \boldsymbol{S^*}^2 \right) ,  \lambda _2=tr\left( \boldsymbol{\varOmega^* }^2 \right) ,  \lambda _3=tr\left( \boldsymbol{S^*}^3 \right) ,  \lambda _4=tr\left( \boldsymbol{\varOmega^* }^2\boldsymbol{S^*} \right) ,  \lambda _5=tr\left( \boldsymbol{\varOmega^* }^2\boldsymbol{S^*}^2 \right). $$

\section{Comparison of the ML and Major Traditional Turbulence Models \label{Ap:B}}
\begin{figure}[!ht]
    \centering
    \includegraphics[width=1\textwidth,trim=0 0 0 35,clip]{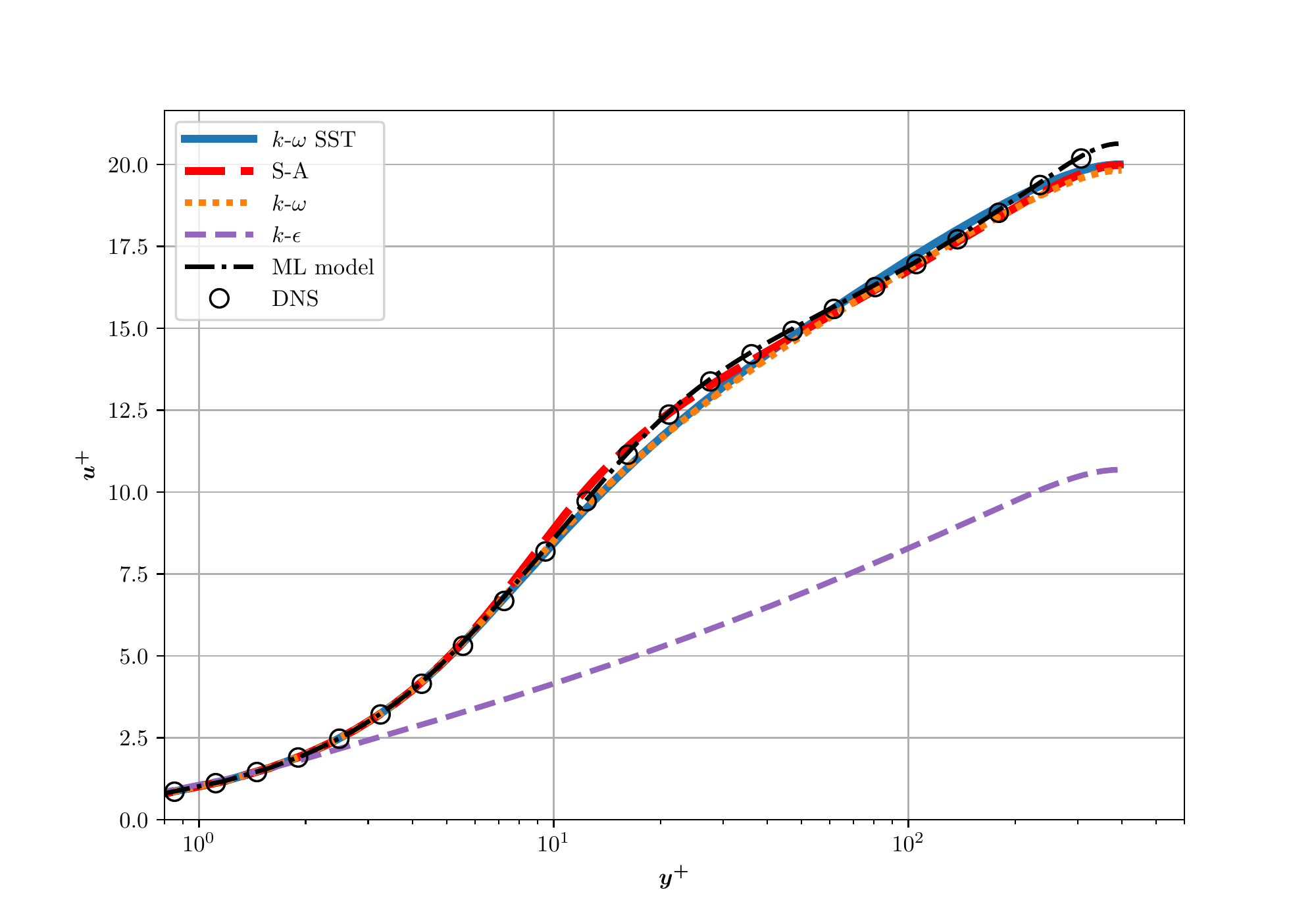}%
    \caption{\label{Fig:comparemodel_upyp} Comparison of velocity profiles in channel flow at $Re_\tau=395$}
\end{figure}

\begin{figure}[!ht]
    \centering
    \includegraphics[width=1\textwidth,trim=55 0 55 0,clip]{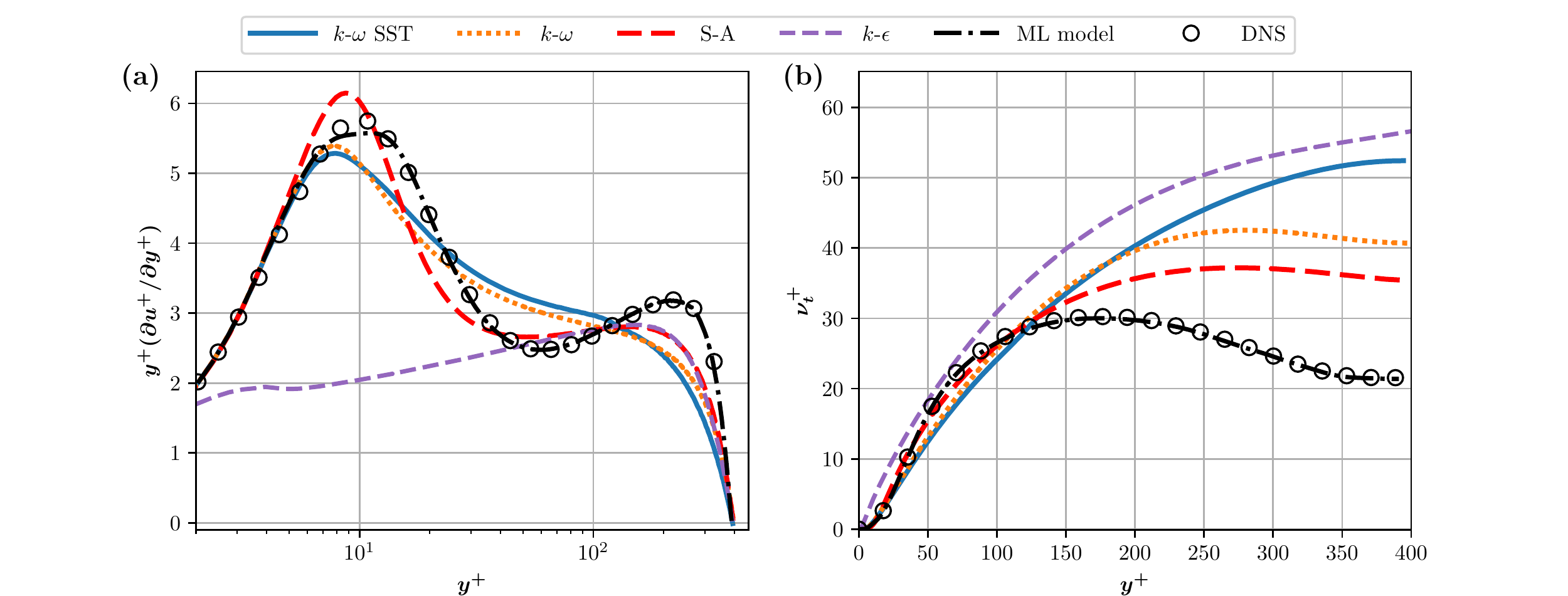}%
    \caption{\label{Fig:comparemodel_2} Comparison of velocity profiles in channel flow at $Re_\tau=395$}
\end{figure}

The profiles of mean velocity, ${u^+}$,  velocity gradient, $y^+(\partial u^+ /\partial y^+ )$, and normalized viscosity, $\nu_t/\nu$, 
in the channel flow at $Re_\tau=395$ from different turbulence models are compared in the Fig.~\ref{Fig:comparemodel_upyp} and Fig.~\ref{Fig:comparemodel_2}, and we observe the ML model presents a superior performance aganist these mainstream turbubulence models, especially for the high-order statistics.

\clearpage

\section*{Acknowledgements}The authors would like to dedicate this paper to Prof. Lipeng Lu, the supervisor of
Weishuo Liu and Jian Fang, who has left us in February 2019. He devoted all his life studying the mystery in
turbulence flow and was always caring for everything about his students beyond academic affairs. This work
would not have been done without his contribution and academic outlook. We will long remember his elegance,
kindness, demeanor, and optimism fighting against fatal illness. The project is supported by the
National Natural Science Foundation of China (Grant Nos. 51420105008). The authors would also like to acknowledge
UKRI-EPSRC under Grant Nos.  EP/R029326/1 and EP/R029369/1 for their financial support. J.F. gratefully acknowledges support by CoSeC, the Computational Science Centre for Research Communities, through UKCTRF.


\bibliography{test}

\end{document}